\documentclass[apj,revtex4]{emulateapj}

\usepackage{natbib,ifthen}
\usepackage{graphicx}
\usepackage{lscape}
\usepackage{amssymb,latexsym}
\usepackage{url}
\usepackage[latin1]{inputenc}
\usepackage{fp}
\usepackage{{aas_macros}}

\slugcomment{Accepted for publication in The Astrophysical Journal 27 April 2015}
\newcommand {\mic} {$\mu$m}
\newcommand {\ben} {\begin{eqnarray}}
\newcommand {\een} {\end{eqnarray}}

\newcommand{\msunperyr}{M$_\odot$ yr$^{-1}$}
\newcommand{\ks}{$K_{\rm s}$}
\newcommand{\jmag}{{\it J}}
\newcommand{\hmag}{{\it H}}

\def\2dust{%
             \leavevmode
             $\raisebox{0.3ex}{\mbox{{\sc 2-D}}}\kern-0.55em%
             \raisebox{-0.3ex}{\mbox{{\sc Dust}}}$}
\def\2dust{{\bf 2-D}{\sc ust}}
\newcommand{\numlmc}{1\,717\,554}
\newcommand{\numsmc}{457\,760}
\newcommand{\numlvars}{2198}
\newcommand{\numsvars}{571}
\newcommand{\numunvarsl}{641}
\newcommand{\numunvarss}{139}

\begin{document}

\title{SAGE-Var: An Infrared Survey of Variability in the Magellanic Clouds}

\author{D. Riebel} 
\affil{Department of Physics, United States Naval Academy, 572C Holloway Road, Annapolis, MD 21402, USA}
\email{riebel.d@gmail.com}

\author{M. L. Boyer}
\affil{Observational Cosmology Laboratory, Code 665, NASA Goddard Space Flight Center, Greenbelt, MD 20771, USA }

\author{S. Srinivasan}
\affil{Academia Sinica, Institute of Astronomy and Astrophysics, P.O. Box 23-141, Taipei 10617, Taiwan}

\author{P. Whitelock}
\affil{Astrophysics, Cosmology and Gravity Centre, Astronomy Department, University of Cape Town, Rondebosch 7701, South Africa}
\affil{South African Astronomical Observatory, PO Box 9, Observatory 7935, South Africa}

\author{M. Meixner}
\affil{Space Telescope Science Institute, 3700 San Martin Drive, Baltimore, MD 21218, USA}
\affil{Department of Physics and Astronomy, The Johns Hopkins University, 3400 N. Charles St. Baltimore, MD 21218, USA}

\author{B. Babler}
\affil{Department of Astronomy, University of Wisconsin-Madison, Madison, WI 53706, USA}

\author{M. Feast}
\affil{Astrophysics, Cosmology and Gravity Centre, Astronomy Department, University of Cape Town, Rondebosch 7701, South Africa}
\affil{South African Astronomical Observatory, PO Box 9, Observatory 7935, South Africa}

\author{M. A. T. Groenewegen}
\affil{Royal Observatory of Belgium, Ringlaan 3, B-1180 Brussels, Belgium}

\author{Y. Ita}
\affil{Astronomical Institute, Graduate School of Science, Tohoku University, 6-3 Aramaki Aoba, Aoba-ku, Sendai, Miyagi 980-8578, Japan}

\author{M. Meade}
\affil{Department of Astronomy, University of Wisconsin-Madison, Madison, WI 53706, USA}

\author{B. Shiao}
\affil{Space Telescope Science Institute, 3700 San Martin Drive, Baltimore, MD 21218, USA}

\author{B. Whitney}
\affil{Department of Astronomy, University of Wisconsin-Madison, Madison, WI 53706, USA}

\shortauthors{Riebel {\it et al.}}
\shorttitle{}

\begin{abstract}
We present the first results from the SAGE-Var program, a follow on to the \textit{Spitzer} legacy program \textit{Surveying the Agents of Galaxy Evolution} \citep[SAGE;][]{Meixner2006}.  We obtained 4 epochs of photometry at 3.6 \&\ 4.5\,\mic\ covering the bar of the Large Magellanic Cloud (LMC) and the central region of the Small Magellanic Cloud (SMC) in order to probe the variability of extremely red sources missed by variability surveys conducted at shorter wavelengths, and to provide additional epochs of observation for known variables.  Our 6 total epochs of observations allow us to probe infrared variability on 15 different timescales ranging from $\sim$20~days to $\sim$5~years.  Out of a full catalog of \numlmc\ (LMC) and \numsmc\ (SMC) objects, we find 10 (LMC) and 6 (SMC) large amplitude AGB variables without optically measured variability owing to circumstellar dust obscuration. The catalog also contains multiple observations of known AGB variables, type I and II Cepheids, eclipsing variables, R CrB stars and young stellar objects which will be discussed in following papers.  Here we present infrared Period-Luminosity (PL) relations for classical Cepheids in the Magellanic Clouds, as well as improved PL relationships for AGB stars pulsating in the fundamental mode using mean magnitudes constructed from 6 epochs of observations.

\end{abstract}

\section{INTRODUCTION}
The study of stellar variability has had a wide-ranging impact on astronomy and cosmology.  Recent variable star surveys such as the MAssive Compact Halo Object search \citep[MACHO;][]{Alcock1997} and the Optical Gravitational Lensing Experiment \citep[OGLE;][]{Udalski1997} have generated catalogs of tens of thousands of variable stars of numerous classes in the Magellanic Clouds.  However, like most ground-based variability surveys, both of these were performed at visible wavelengths, and therefore miss the reddest variable sources, such as dust-enshrouded, highly evolved Asymptotic Giant Branch (AGB) stars, which are nearly invisible except in the infrared (IR).  AGB stars are unstable, and exhibit variability on timescales of hundreds of days \citep{Vassiliadis1993}.  Due to this, they are also classified as Long Period Variables (LPVs).  There are several surveys that monitored LPVs at near-IR wavelengths in the Magellanic Clouds \citep[$J$, $H$, and $K$-bands;][]{Ita2002,Whitelock2003}, but even these can still miss the dustiest sources. The pulsation properties of the heavily enshrouded (and most evolved) AGB stars are therefore not well understood. This problem can be addressed with IR monitoring at wavelengths longer than 3\,\mic, but there are very few examples of mid-IR monitoring surveys in any galaxy. \citet{Le Bertre1992,Le Bertre1993} obtained light curves at 1--20\,\mic\ of $\sim$60 O- and C-rich AGB stars in our own galaxy and found that while pulsation amplitudes generally decrease into the mid-IR, circumstellar dust can cause amplitudes to increase again at $\lambda > 3$\,\mic. In M33, \citet{McQuinn2007} obtained 5 epochs of imaging at 3.6, 4.5, 5.8, and 8\,\mic\ and found that the pulsation amplitude tends to increase with color, but that this relationship may break down at the longest wavelengths. In the Magellanic Clouds and other nearby dwarf galaxies, only 2--3 epochs of imaging is available at $\lambda > 3$\,\mic\ \citep{Polsdofer2015,Vijh2009, Boyer2015a,Boyer2015b}. While these surveys can detect flux changes in a large fraction of the dustiest stars, they cannot place stringent limits on the pulsation periods or amplitudes in the infrared.  In order to explore the variability of these reddest sources, which dominate the mass return to the interstellar medium (ISM) from evolved stars \citep{Riebel2012}, we used the \textit{Spitzer Space Telescope} to survey the bar regions of the LMC and SMC at 3.6 and 4.5\,\mic.  SAGE-Var represents the first large scale variability survey at such red wavelengths.  While our original focus was the reddest AGB stars, we have detected over 2,700 IR variables in the Clouds of many classes.

This paper is organized as follows:  In \S\,\ref{sec:var_data} we detail the observational strategy (\ref{sec:obs}), data reduction (\ref{sec:reduction}), catalog construction (\ref{sec:catalog}), and source classification scheme (\ref{sec:source_class}) for the SAGE-Var program.  In \S\,\ref{sec:results}, we discuss our results; the identification of Long-Period Variable (LPV) candidates without prior observed variability (\ref{sec:new_lpvs}), an investigation of the connection between variability amplitude and Dust Production Rate (DPR) from evolved stars (\ref{sec:amp_dpr}), and new measurements of the IR Period Luminosity relationship for both LPVs and classical Cepheids in the Magellanic Clouds (\ref{sec:pl}).  Our conclusions are presented in \S\,\ref{sec:conclusions}.

\section{THE DATA} \label{sec:var_data}
\subsection{Observations} \label{sec:obs}
\begin{figure}
\begin{center}
\rotatebox{0}{\includegraphics*[width=0.65\linewidth]{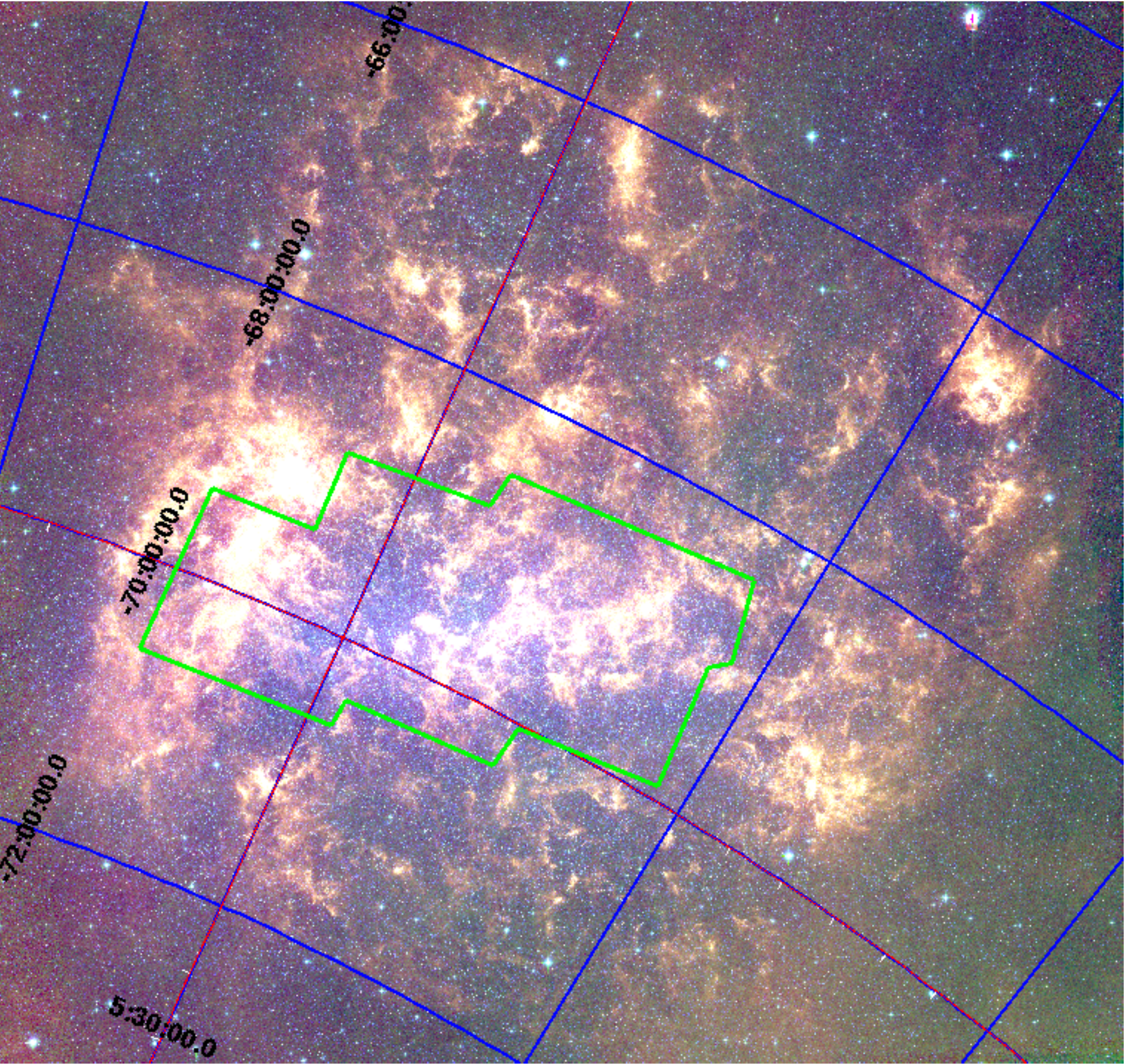}}
\caption[Footprint of the SAGE-Var observations in the LMC]{Three-color image of the LMC (red: 8.0\,\mic, green: 5.8\,\mic, blue: 3.6\,\mic) showing the footprint of the SAGE-Var observations in green, focused on the stellar bar region of the galaxy.}
\label{fig:lmc_footprint}
\end{center}
\end{figure}

\begin{figure}
\begin{center}
\rotatebox{0}{\includegraphics*[width=0.65\linewidth]{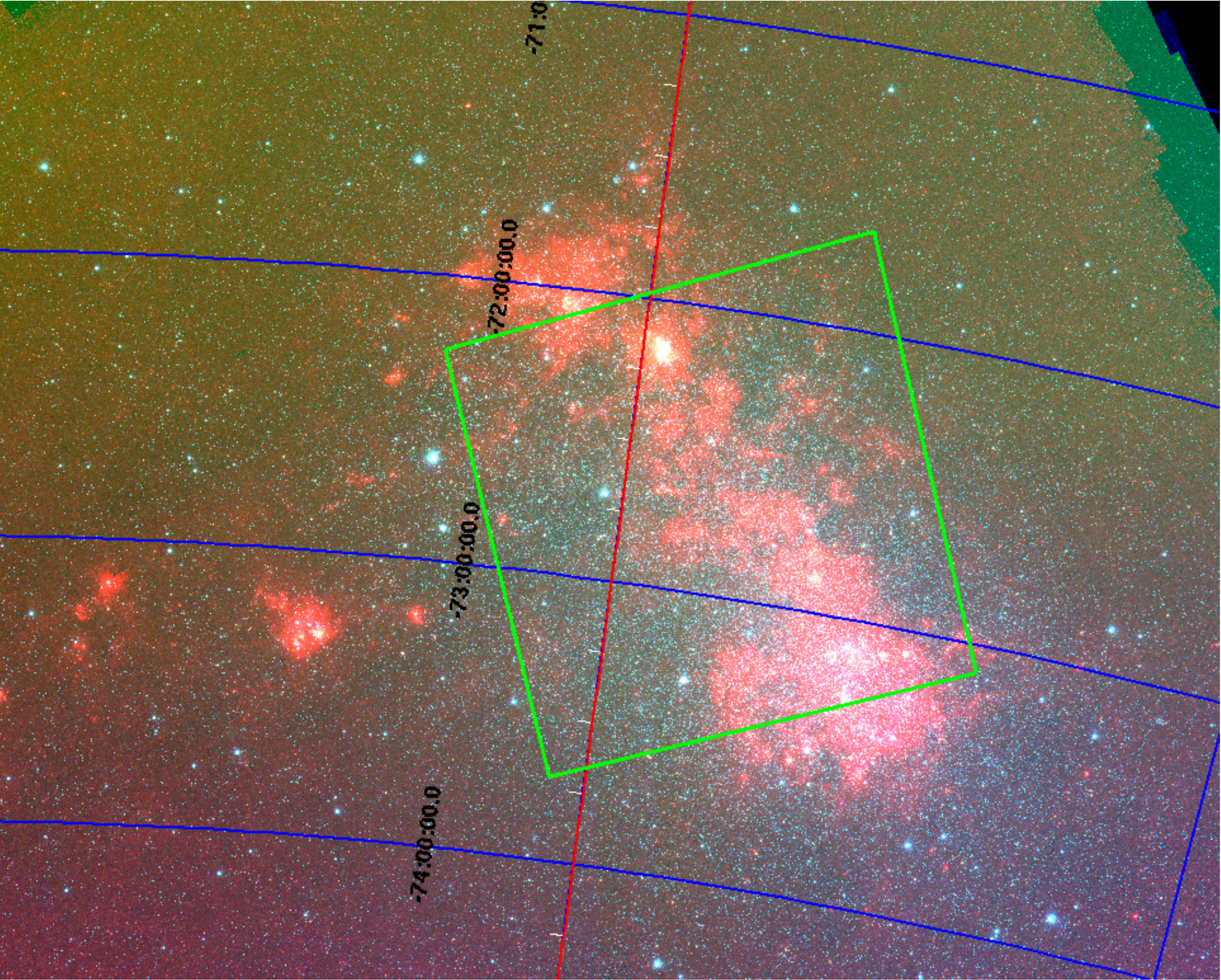}}
\end{center}
\caption[Footprint of the SAGE-Var observations in the SMC]{Three-color image of the SMC (red: 24\,\mic, green: 4.5\,\mic, blue: 3.6\,\mic) showing the footprint of the SAGE-Var observations in green, focused on the main stellar locus of the galaxy.}
\label{fig:smc_footprint}
\end{figure}

The 4 epochs of SAGE-Var observations ({\it Spitzer} PID 70020) were taken over a 10 month period, between August 2010 and June 2011 (see Table~\ref{tab:obs_date}).  Each epoch is a $3.7^{\circ} \times 1.5^{\circ}$ (LMC, Figure~\ref{fig:lmc_footprint}) or $1.7^{\circ} \times 1.7^{\circ}$ (SMC, Figure~\ref{fig:smc_footprint}) mapping of the bar region of the galaxy, using both the [3.6] and [4.5] bands of the \textit{Spitzer} warm mission.  All frames were taken as 12\,second exposures using the IRAC High Dynamic Range (HDR) mode, which also produces short 0.6\,second exposures in order to mitigate the effects of saturation for the brightest sources.  This exposure mode was chosen to provide uniformity with the previous two epochs of the SAGE-LMC \citep{Meixner2006} and SAGE-SMC \citep{Gordon2011} surveys.  The scheduling of our observations was determined using a \citet{Madore2005} power-law cadence with an index of 0.99, taking into account the original two SAGE epochs.  Including these two original epochs, our 6 total epochs of observations allow us to probe variability on 15 different timescales (see~\S\,\ref{sec:catalog}).

\begin{deluxetable}{lcc}
\tabletypesize{}
\tablecaption{Observation dates}
\tablewidth{0pt}
\tablehead{\colhead{Epoch} & \colhead{Date} & \colhead{Julian Date} }
\startdata
& LMC:  & \\
Epoch 1: & 2005 Jul 20 & 2453572 \\
Epoch 2: & 2005 Oct 28 & 2453672 \\
Epoch 3: & 2010 Aug 17 & 2455426 \\
Epoch 4: & 2010 Sep 10 & 2455450 \\
Epoch 5: & 2010 Dec 25 & 2455556 \\
Epoch 6: & 2011 Apr 27 & 2455679 \\
& SMC:  & \\
Epoch 1: & 2008 Jun 15 & 2454633 \\
Epoch 2: & 2008 Sep 19& 2454729 \\
Epoch 3: & 2010 Aug 17 & 2455426 \\
Epoch 4: & 2010 Sep 12 & 2455452 \\
Epoch 5: & 2010 Dec 24 & 2455555 \\
Epoch 6: & 2011 Jun 16 & 2455729
\enddata
\tablecomments{Dates of the 6 epochs of observations of the SAGE-Var project, including the 2 epochs of the original SAGE-LMC and SAGE-SMC programs (Epochs 1 \& 2).  The dates listed for the original two epochs of observations are the approximate midpoints of the $\sim$5~day observation periods, while later epochs were executed during a single 24-hour period.}
\label{tab:obs_date}
\end{deluxetable}

\subsection{Data Reduction} \label{sec:reduction}
The SAGE-Var data were processed using a pipeline developed at the University of Wisconsin for the GLIMPSE survey \citep{Benjamin2003}, which was also used for processing the original SAGE-LMC and SAGE-SMC observations.  The Wisconsin pipeline corrects for numerous observational artifacts such as stray light, column pulldown, banding, and bad pixels.  These observational artifacts are discussed by \citet{Hora2004} and the Wisconsin pipeline is described in section 3.1.2 of \citet{Meixner2006}.  The pipeline is based on DAOPHOT \citep{Stetson1987} and is modified to account for the variable diffuse background in IRAC images.  Photometry is performed on the Basic Calibrated Data (BCD) frames, and the FWHM of the point-spread function is $\sim$1.7\arcsec\ in both bands.

 In addition, the individual frames are mosaicked, and point sources identified.  During catalog construction (\S~\ref{sec:catalog}), the point source lists are also position matched to 2MASS \jmag \hmag \ks\ photometry.  The 2MASS bands have been dereddened to account for interstellar extinction.  The reddening coefficients used for the LMC can be found in Table~1 of \citet{Riebel2012}.  The photometry for the more distant SMC was dereddened using a value of $E_{B-V}=0.04$~mag \citep{Harris2004, Schlegel1998} and the prescription of \citet[][pp.109--111]{Glass1999}.  Our fluxes were transformed into magnitudes using zero points of 6.12\,mag (280.9\,Jy) for [3.6] and 5.63\,mag (179.7\,Jy) for [4.5] \citep{Reach2005}.

\begin{figure*}
\centerline{
\begin{tabular}{cc}
\rotatebox{270}{\includegraphics*[scale=0.4]{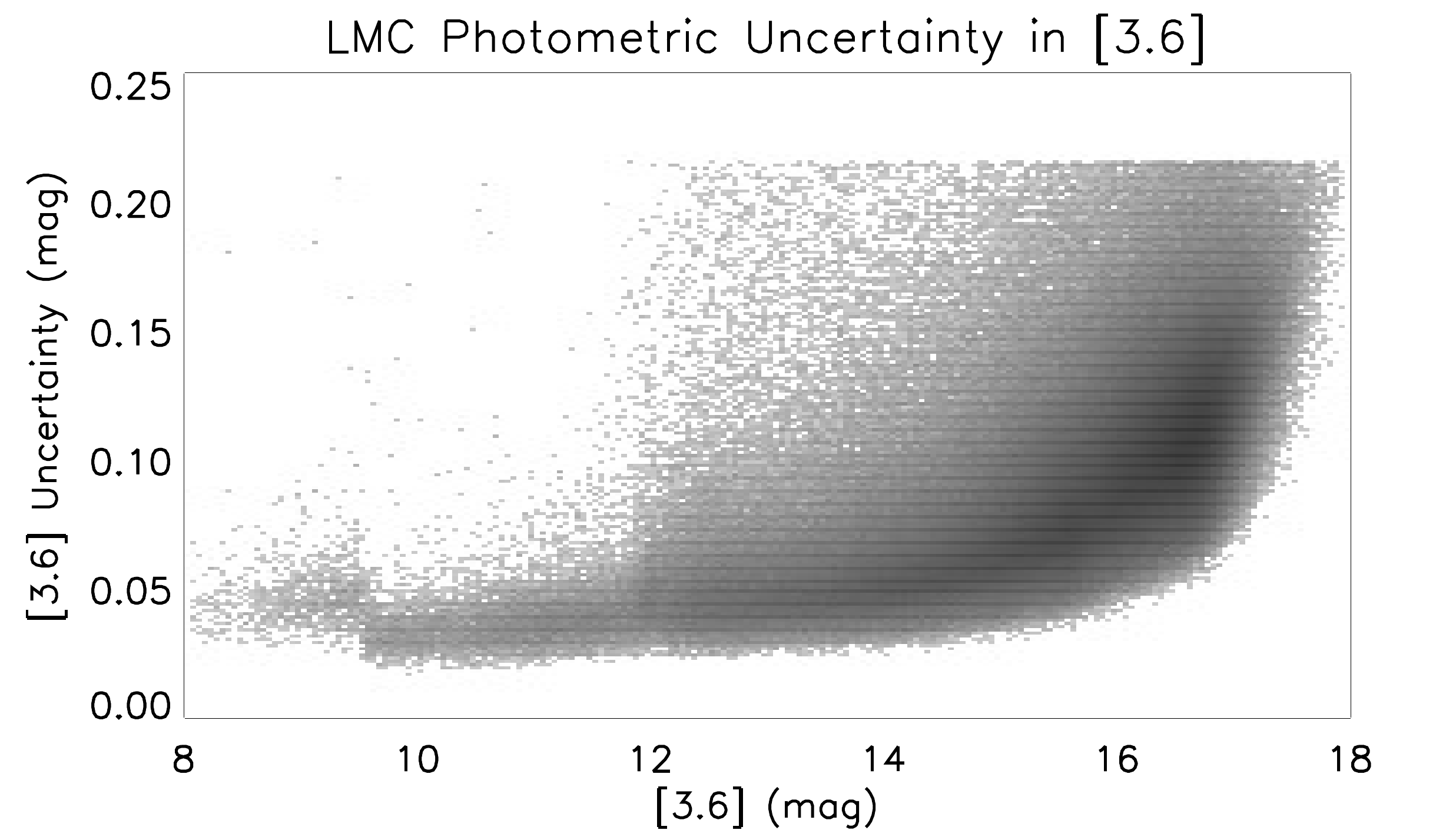}} & \rotatebox{270}{\includegraphics*[scale=0.40]{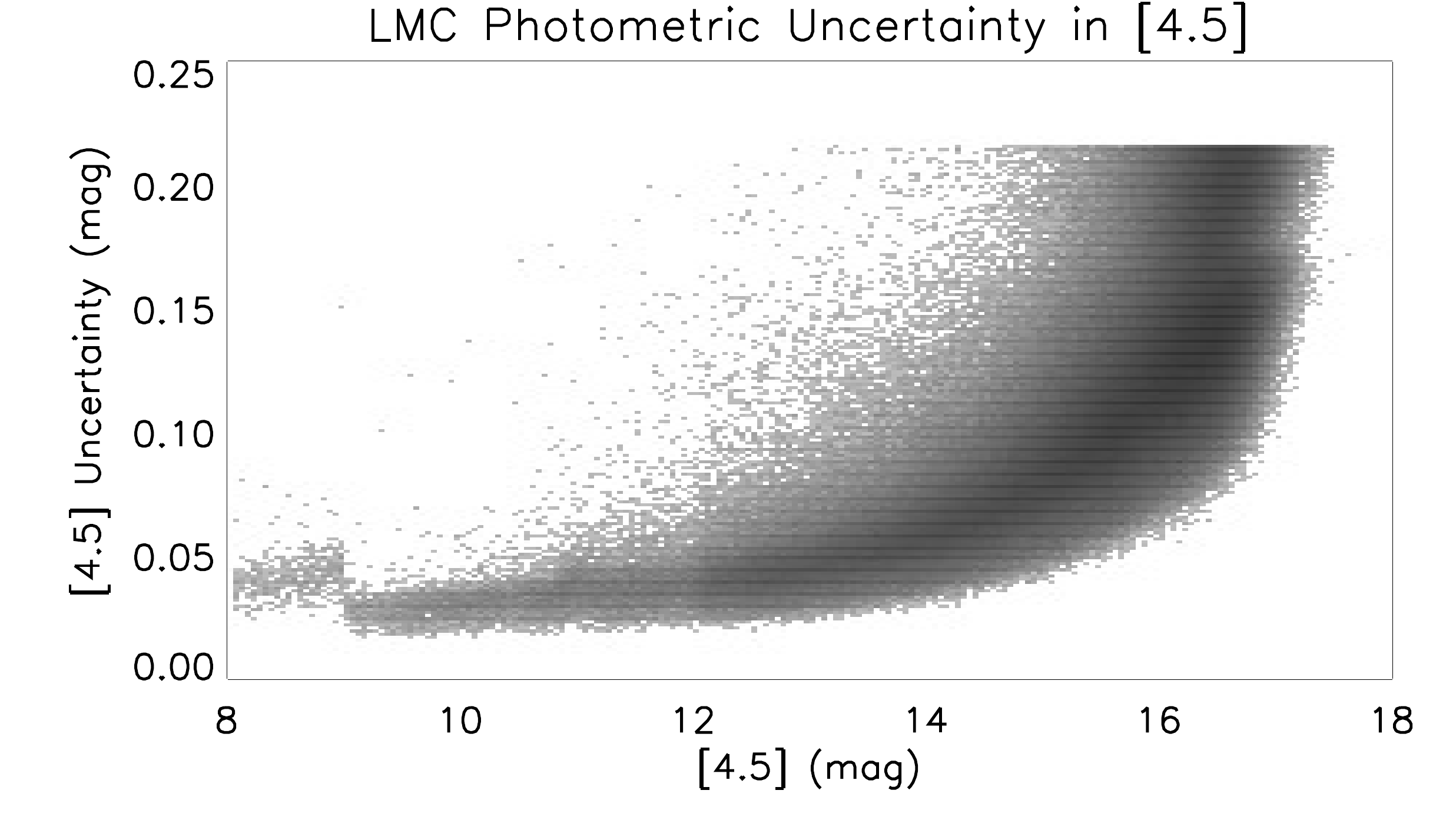}} \\
 \rotatebox{270}{\includegraphics*[scale=0.4]{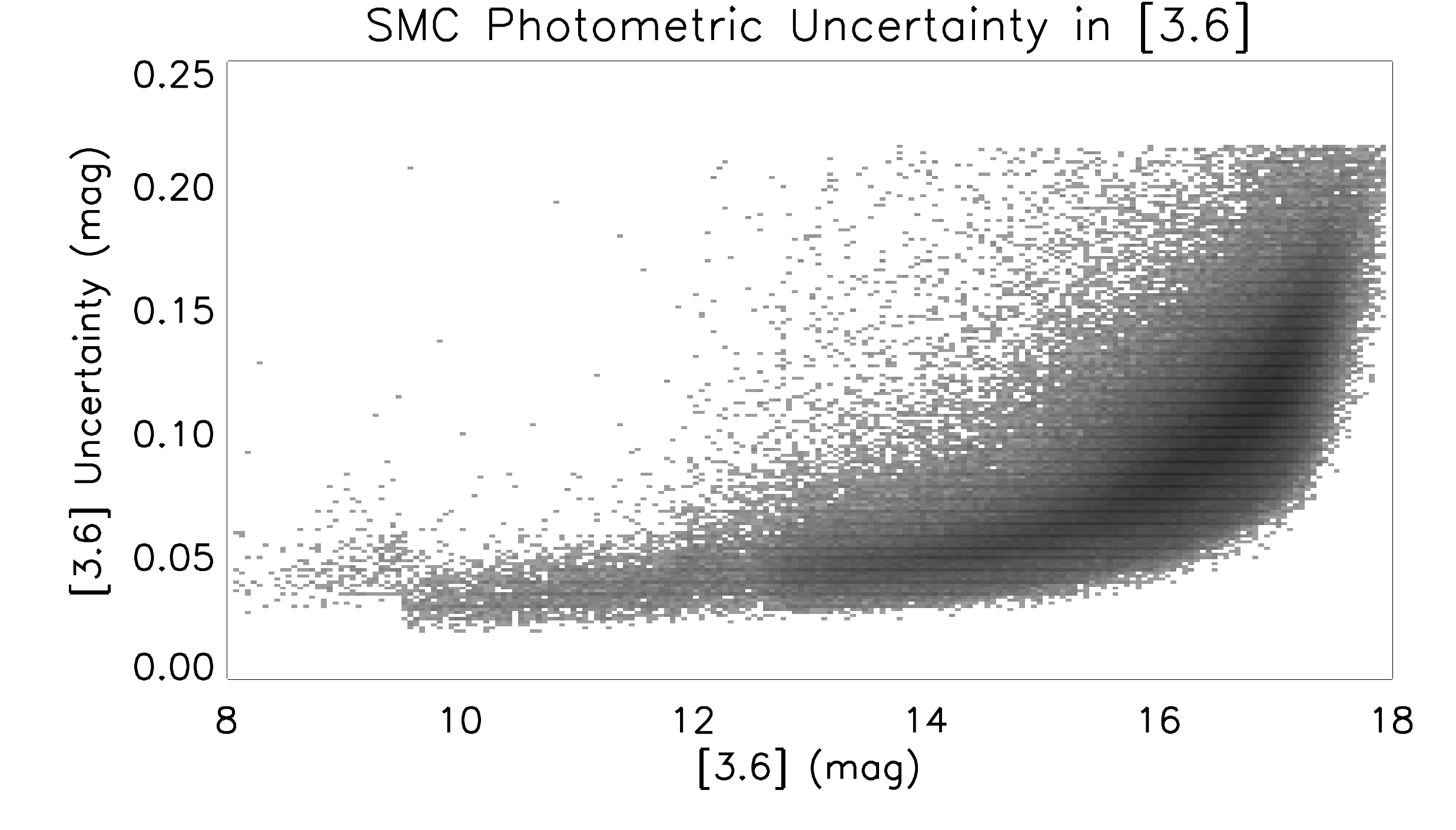}} & \rotatebox{270}{\includegraphics*[scale=0.40]{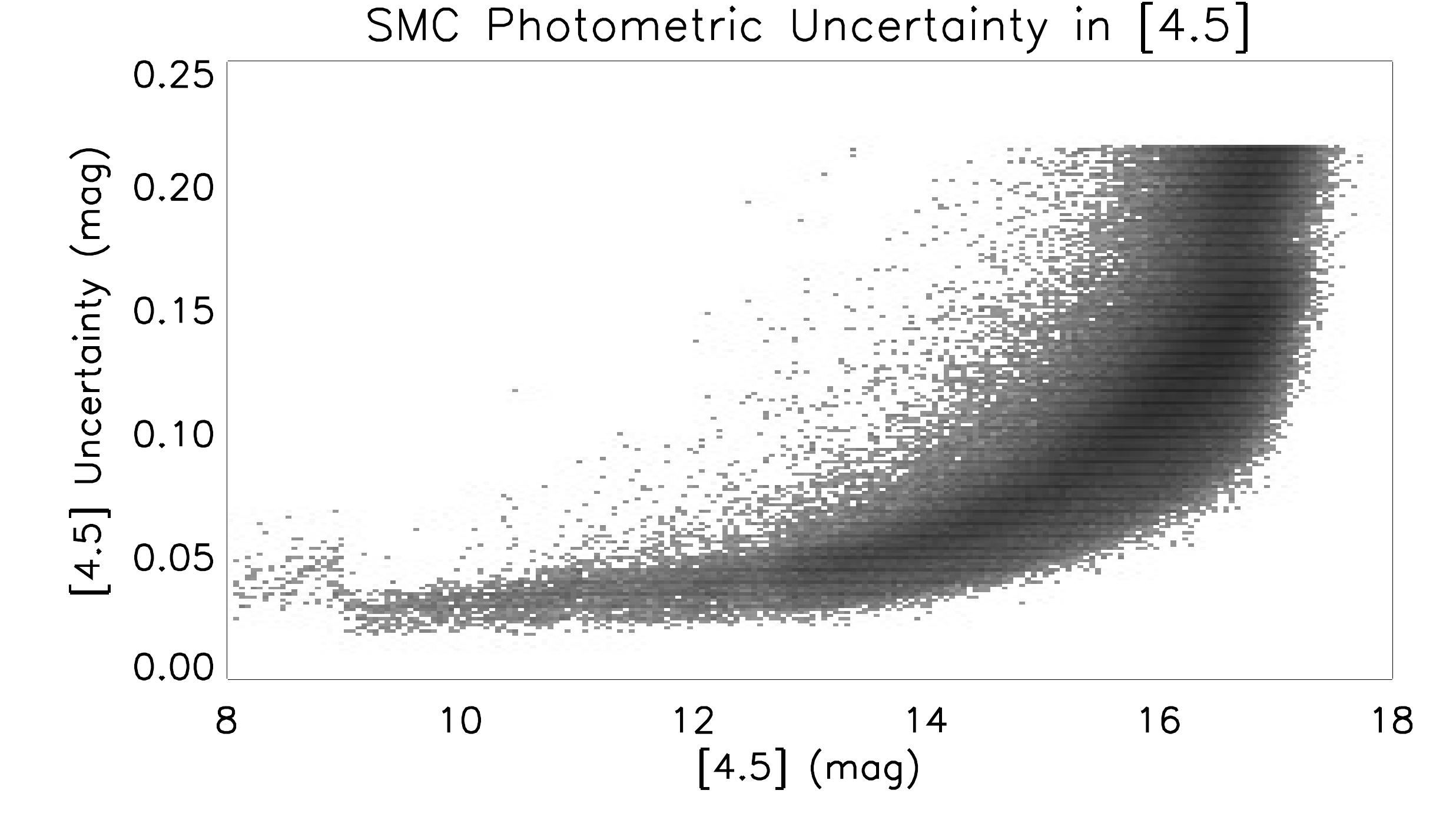}}
\end{tabular}
}
\caption[Photometric Uncertainty as a function of source magnitude]{Each plot shows the 1$\sigma$ uncertainty produced from the Wisconsin pipeline as a function of source magnitude for the entire SAGE-Var dataset as a Hess diagram.  The top row shows the data from the LMC ([3.6] on the left, [4.5] on the right), while the bottom row shows the SMC data.}
\label{fig:photo_uncert}
\end{figure*}

The pipeline provides 1\,$\sigma$ photometric uncertainty in each band, plotted as a function of source magnitude in Figure~\ref{fig:photo_uncert}.  Throughout this paper, we take these 1\,$\sigma$ errors as the uncertainties in our photometry.  The distributions in Fig.~\ref{fig:photo_uncert} show a hard cutoff at 0.2 mag due to the requirement that sources in the original SAGE survey have a S/N $>$ 5, which effectively results in the top cutoff.    Sources brighter than magnitude $\sim$9.0 are observed only using the 0.6\,s exposure mode, which results in a larger uncertainty than the 12\,s exposure time.  Between magnitudes 9 and 12, the 0.6\,s and 12\,s exposures are averaged, and the photometry of sources dimmer than magnitude $\sim$12 is derived only from the 12\,s exposures.  These three regimes of source brightness result in the breaks visible at $\sim$9 and $\sim$12 mag.  See \S\,3.1.3 of \citet{Meixner2006} for a full discussion.

\subsection{Catalog Construction} \label{sec:catalog}
The initial data product was a `full' point source list, without cosmic ray screening.  These artifacts were removed from the catalog by matching the full lists to the SAGE mosaic photometry archive\footnote{\url{http://irsa.ipac.caltech.edu/data/SPITZER/SAGE/doc/ \\ SAGEDataProductsDescription\_Sep09.pdf}} using a 2\arcsec\  matching radius.  The mosaic archives of the original SAGE surveys were constructed by co-adding and re-reducing the photometry from the original two SAGE Epochs.  The archives are deeper, more complete, and of higher signal-to-noise (S/N) than the SAGE-Var data due to the greater exposure time of the mosaic photometry (up to $\sim$50\,s per pixel compared to 12\,s) and the fact that the original surveys were performed during the cold phase of the \textit{Spitzer} mission whereas the SAGE-Var observations were taken after the liquid helium cryogen aboard \textit{Spitzer} was exhausted.  The higher operating temperature of the \textit{Spitzer} instruments restricts our observations to only the shortest IRAC wavelengths, and lends additional thermal noise to our images.  Due to their greater depth and quality, the SAGE mosaic archives serve as a `truth field' for the SAGE-Var full source lists, providing a more thorough list of actual astronomical sources while screening out instrumental artifacts by position matching.  After the SAGE-Var source list was matched to the mosaic photometry, the original SAGE Epoch 1 and 2 archive data were matched individually as well, producing a final source list of \numlmc\ (LMC) and \numsmc\ (SMC) unique sources with up to 6 individual 3.6 and 4.5\,\mic\ observations, as well as mosaic photometry available in the longer wavelength IRAC bands.  Table~\ref{tab:var_pops} lists the number of unique sources detected in each band of each epoch of SAGE-Var observations.  Also listed is the number of objects detected in only one epoch, in two epochs, in three epochs, etc.  A greater number of detections is preferable.  The entire SAGE-Var data set, including all 6 epochs of observations in both bands, is available as an online table, hosted at the NASA/IPAC Infrared Science Archive\footnote{\url{http://irsa.ipac.caltech.edu/Missions/spitzer.html}}.

\begin{deluxetable}{lcccc}
\tabletypesize{}
\tablecaption{Object Count}
\tablecomments{Summary of the number of unique objects detected in the SAGE-Var survey of the SMC and LMC}
\tablewidth{0pt}
\tablehead{ \colhead{Epoch} & \multicolumn{2}{c}{LMC} &  \multicolumn{2}{c}{SMC} \\
 \colhead{} & \colhead{3.6\,\mic} & \colhead{4.5\,\mic} & \colhead{3.6\,\mic} & \colhead{4.5\,\mic} }
\startdata
Mosaic Photometry &      1,712,135 &      1,712,994 &       454,951 &       455,587\\
SAGE Epoch 1 &      1,185,774 &      1,083,380 &       297,732 &       269,178\\
SAGE Epoch 2 &      1,163,888 &      1,068,438 &       290,055 &       262,408\\
SAGE-Var 1 &      1,177,715 &       824,928 &       289,359 &       185,392\\
SAGE-Var 2 &      1,242,129 &       852,734 &       264,921 &       176,911\\
SAGE-Var 3 &      1,244,543 &       855,533 &       287,168 &       173,756\\
SAGE-Var 4 &      1,244,791 &       845,370 &       280,656 &       172,379\\
1 Detection &       116,285 &       240,980 &        56,602 &        78,408\\
2 Detections &       197,701 &       240,983 &        75,528 &        61,853\\
3 Detections &       270,324 &       273,872 &        79,241 &        65,429\\
4 Detections &       253,951 &       183,972 &        54,222 &        45,670\\
5 Detections &       271,603 &       180,249 &        64,538 &        43,291\\
6 Detections &       593,727 &       391,448 &       120,822 &        73,748\\
Total Sources& \multicolumn{2}{c}{1,717,554\tablenotemark{a}} & \multicolumn{2}{c}{457,760\tablenotemark{a}}
\enddata
\label{tab:var_pops}
\tablenotetext{a}{These numbers represent the total number of unique sources detected in each galaxy, with an absolute minimum of one detection in one band.}
\end{deluxetable}

\begin{deluxetable*}{cllc}
\tabletypesize{\scriptsize}
\tablecaption{Full Object Catalog Contents}
\tablewidth{0pt}
\tablehead{ 
\colhead{Column} &  \colhead{Name} & \colhead{Description} & \colhead{Null}
}
\startdata
1 &  desig & Source IRAC Designation & \nodata  \\
2 & ra & Right Ascension, J2000 [deg] & \nodata \\
3 & dec & Declination, J2000 [deg] & \nodata \\
4&  e1\_36 & Epoch 1 flux in [3.6] band [Jy] & $-99$ \\
5&  e1\_36\_u & Uncertainty in Epoch 1 [3.6] flux [Jy] & $-99$ \\
6--15& eN\_36 &Epoch 2--6 flux/uncertainty in [3.6] band\tablenotemark{a} [Jy] & $-99$ \\
16--27& eN\_45 &Epoch 1--6 flux/uncertainty in [4.5] band\tablenotemark{b} [Jy] & $-99$ \\
28 &  mean\_36&Mean [3.6] flux [Jy] & $-99$ \\
29 &  mean\_36\_u&RMS Uncertainty in mean [3.6] flux [Jy] & $-99$ \\
30 &  mean\_45&Mean [4.5] flux [Jy] & $-99$ \\
31 & mean\_45\_u& RMS Uncertainty in mean [4.5] flux [Jy] & $-99$ \\
32--46 & var\_36\_N & Variability index at [3.6] for interval\tablenotemark{c} N & NaN \\
47--61 & var\_45\_N & Variability index at [4.5] for interval\tablenotemark{c} N & NaN \\
62& ogle\_id & ID From the OGLE-III Catalog of Variable Stars & \nodata \\
63&ogle\_class & Classification from the OGLE-III CVS & \nodata \\
64& ogle\_per & Variability period from the OGLE-III CVS [days] & $-99$ \\
65& macho\_id & ID from the MACHO survey & \nodata \\
66& macho\_per& Variability period from the MACHO survey [days] & $-99$ \\
67&grams\_class&Classification of best fitting GRAMS model (C or O) & \nodata \\
68&yso\_class & `Y' if a source is classified as a YSO candidate & \nodata \\
\enddata
\tablecomments{The full SAGE-Var catalog of \numlmc\  (\numsmc) sources in the LMC (SMC) is available from IRSA.  This table is provided as a guide to the online catalog's structure and content.}
\tablenotetext{a}{The epoch 2 through 6 [3.6] photometry follows the same format as columns 4 and 5}
\tablenotetext{b}{The [4.5] photometry follows the same format as the [3.6] photometry in columns 4-15}
\tablenotetext{c}{Defined in  \S~\ref{sec:source_class} and Table~\ref{tab:interval}}
\label{tab:full_catalog}
\end{deluxetable*}

\begin{deluxetable*}{cllc}
\tabletypesize{\scriptsize}
\tablecaption{Variable Object Catalog Contents}
\tablewidth{0pt}
\tablehead{ 
\colhead{Column} &  \colhead{Name} & \colhead{Description} & \colhead{Null}
}
\startdata
1 &  desig & Source IRAC Designation & \nodata  \\
2 & ra & Right Ascension, J2000 [deg] & \nodata \\
3 & dec & Declination, J2000 [deg] & \nodata \\
4&  e1\_36 & Epoch 1 flux in [3.6] band [Jy] & $-99$ \\
5&  e1\_36\_u & Uncertainty in Epoch 1 [3.6] flux [Jy] & $-99$ \\
6--15& eN\_36 &Epoch 2--6 flux/uncertainty in [3.6] band\tablenotemark{a} [Jy] & $-99$ \\
16--27& eN\_45 &Epoch 1--6 flux/uncertainty in [4.5] band\tablenotemark{b} [Jy] & $-99$ \\
28 &  mean\_36&Mean [3.6] flux [Jy] & $-99$ \\
29 &  mean\_36\_u&RMS Uncertainty in mean [3.6] flux [Jy] & $-99$ \\
30 &  mean\_45&Mean [4.5] flux [Jy] & $-99$ \\
31 & mean\_45\_u& RMS Uncertainty in mean [4.5] flux [Jy] & $-99$ \\
32--46 & var\_36\_N & Variability index at [3.6] for interval\tablenotemark{c} N & NaN \\
47--61 & var\_45\_N & Variability index at [4.5] for interval\tablenotemark{c} N & NaN \\
62& ogle\_id & ID From the OGLE-III Catalog of Variable Stars & \nodata \\
63&ogle\_class & Classification from the OGLE-III CVS & \nodata \\
64& ogle\_per & Variability period from the OGLE-III CVS [days] & $-99$ \\
65& macho\_id & ID from the MACHO survey & \nodata \\
66& macho\_per& Variability period from the MACHO survey [days] & $-99$ \\
67&grams\_class&Classification of best fitting GRAMS model (C or O) & \nodata \\
68&yso\_class & `Y' if a source is classified as a YSO candidate & \nodata \\
69&amp\_36 & SAGE-Var observed [3.6] Amplitude & \nodata \\
70 & amp\_45 & SAGE-Var observed [4.5] Amplitude & \nodata 
\enddata
\tablecomments{This table extracts just those sources flagged as variable by the criteria of \S~\ref{sec:source_class}.  It follows essentially the same format as Table~\ref{tab:full_catalog} with the addition of the observed amplitudes of the variable objects.  These are simply the difference between the brightest and dimmest magnitudes observed for the source, and represent a lower limit on the source's full variability.}
\tablenotetext{a}{The epoch 2 through 6 [3.6] photometry follows the same format as columns 4 and 5}
\tablenotetext{b}{The [4.5] photometry follows the same format as the [3.6] photometry in columns 4-15}
\tablenotetext{c}{Defined in  \S~\ref{sec:source_class} and Table~\ref{tab:interval}}
\label{tab:var_catalog}
\end{deluxetable*}

\subsection{Source Classification} \label{sec:source_class}
Variables in the SAGE-Var dataset were identified using the variability criteria of \citet{Vijh2009}.  We calculate variability indices $$V_{ij}^b = \frac{f_i^b - f_j^b}{\sqrt{(\sigma_i^b)^2 + (\sigma_j^b)^2}}$$ for every star, for every possible combination of epochs $i$ and $j$, for the flux $f$ in each SAGE-Var band $b$, ([3.6] and [4.5]) with photometric uncertainty $\sigma$.  The photometric uncertainties are taken directly from the Wisconsin pipeline (\S\,\ref{sec:reduction}).  The variability index is thus the number of standard errors by which two epochs differ in brightness.  \citet{Vijh2009} had only the first two epochs of the original SAGE survey, but 5 bands of photometry.  In this study, our situation is reversed, in that we have only two bands of photometry, but ${6\choose 2} = 15$ possible epochal differences.  Each epochal difference probes variability on a different timescale (Table~\ref{tab:interval}), and we term each epochal difference an \textit{interval}.  In order for a source to be classified as a variable in a given interval, we require it to exhibit 3$\sigma$ flux variation in the same direction (that is, brightening or dimming) in both the [3.6] and [4.5] bands ($|V_{ij}^b| \geq 3$ and the same sign in both bands).

The SAGE-Var sample has a total of 819 sources in common with the sample of \citet{Vijh2009}.  Of these, we independently identify 752 (92\%) as variables using our own criteria.  The remaining 67 sources typically show marginally variable behavior, with variability indices very close to, but not quite exceeding, our 3$\sigma$ level.  We manually add to our catalog 66 of these 67 sources, omitting one source only detected in one of the four new epochs of SAGE-Var.

\begin{deluxetable}{lccc}
\tabletypesize{}
\tablecaption{Interval Timescales}
\tablewidth{0pt}
\tablehead{\colhead{Epochs} & \colhead{LMC Interval} & \colhead{SMC Interval} \\
 & \colhead{(days)} & \colhead{(days)} & \colhead{label}}
\startdata
Epoch  1 $-$ Epoch  2 &  100  & 96 & 1\\
Epoch  1 $-$ Epoch  3 & 1854  & 793&  2\\
Epoch  1 $-$ Epoch  4 & 1878  & 819 & 3\\
Epoch  1 $-$ Epoch  5 & 1984  & 922 & 4\\
Epoch  1 $-$ Epoch  6 & 2107 & 1096 & 5\\
Epoch  2 $-$ Epoch  3 & 1754  & 697 & 6\\
Epoch  2 $-$ Epoch  4 & 1778  & 723 & 7\\
Epoch  2 $-$ Epoch  5 & 1884  & 826 & 8\\
Epoch  2 $-$ Epoch  6 & 2007  & 1000 & 9 \\
Epoch  3 $-$ Epoch  4 &   24  & 26 & 10\\
Epoch  3 $-$ Epoch  5 &  130  & 129 & 11\\
Epoch  3 $-$ Epoch  6 &  253  & 303 & 12\\
Epoch  4 $-$ Epoch  5 &  106  & 103 & 13\\
Epoch  4 $-$ Epoch  6 &  229  & 277 & 14\\
Epoch  5 $-$ Epoch  6 &  123  & 174& 15
\enddata
\tablecomments{By taking the difference between all possible combinations of observation epochs, we probe source variability on 15 timescales, from $\sim$1\,month to $\sim$5.5\,years.  Throughout this paper, we refer to these epochal differences as intervals.  The ``label" column refers to the electronic table of the entire SAGE-Var catalog, available from the NASA/IPAC Infrared Science Archive (IRSA) (columns 32-61, Tables~\ref{tab:full_catalog} \& \ref{tab:var_catalog}).}
\label{tab:interval}
\end{deluxetable}

\begin{figure}
\begin{center}
\centerline{
\begin{tabular}{c}
\rotatebox{270}{\includegraphics*[width=0.65\linewidth]{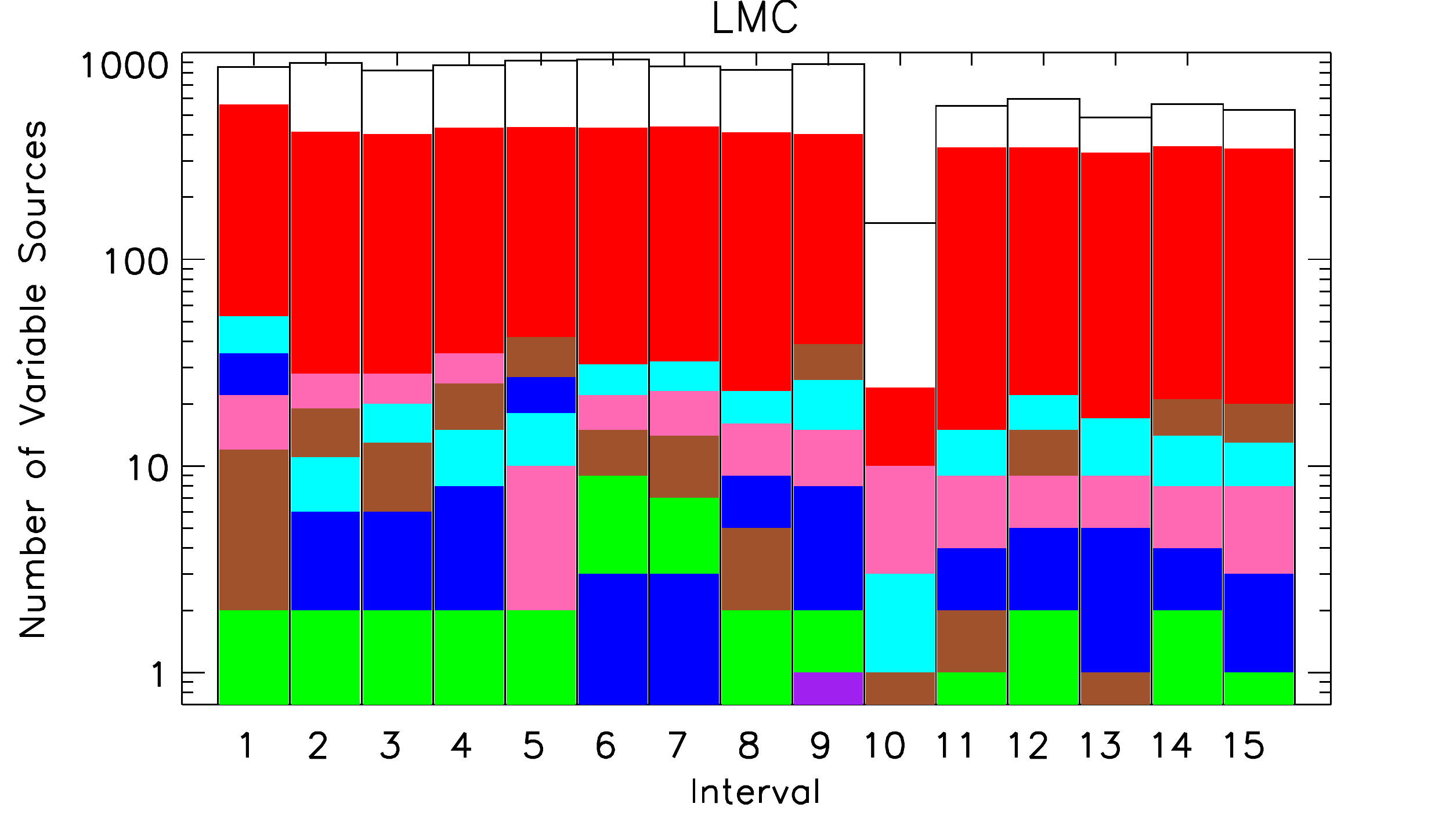}} \\
\rotatebox{270}{\includegraphics*[width=0.65\linewidth]{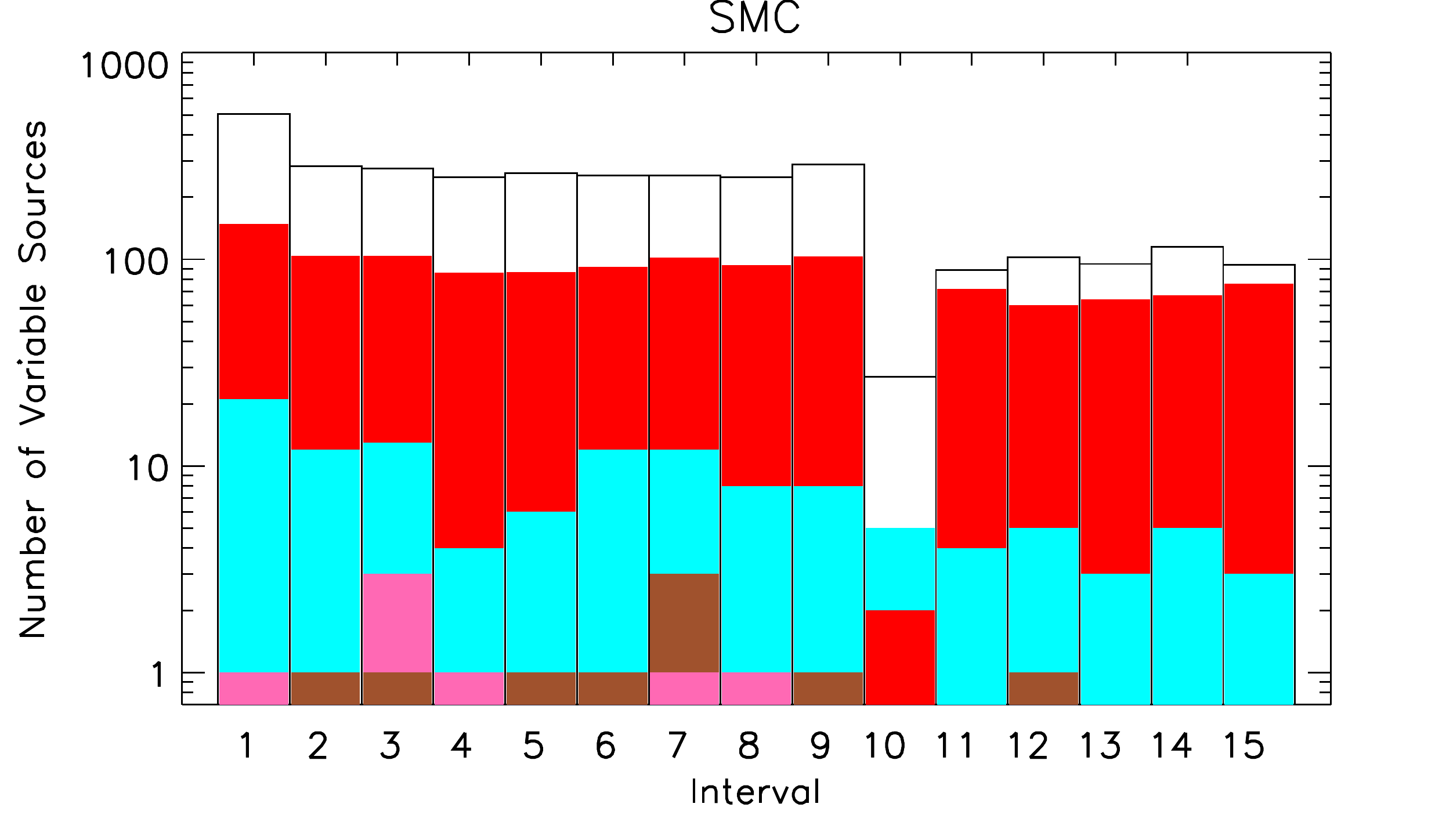}}
\end{tabular}
}
\caption[Logarithmic histogram of number of variables detected in each interval of SAGE-Var]{Histogram of the number of sources detected as variable in each interval of SAGE-Var.  The top panel shows the source distribution in the LMC, while the bottom panel is that for the SMC.  The colored portion of each bar represents the number of sources classified by the OGLE project.  Red: LPVs, Cyan: Classical Cepheids, Hot Pink: Type II Cepheids, Blue: Eclipsing Binaries, Brown: YSOs, Green: R CrB stars, Purple: RR Lyrae stars.  The white area at the top of each interval represents the unclassified variables seen in SAGE-Var but not listed in the OGLE-CVS (Figure~\ref{fig:var_cmd_notogle}).  The labels for the intervals on the $x$-axis refer to the indices in Table~\ref{tab:interval}.  Classes are plotted from least numerous to most numerous, bottom to top.}
\label{fig:var_hist}
\end{center}
\end{figure}

These criteria resulted in \numlvars\ unique variables in the LMC along with \numsvars\ in the SMC.  Histograms showing the number of sources seen as variable in each interval are shown in Figure~\ref{fig:var_hist}.  Intervals 1--9 show a generally higher number of detected variables because they all compare at least one of the initial two SAGE epochs to later observations.  Taken during the cold \textit{Spitzer} mission, which had a S/N approximately twice that obtained during the warm mission when all other epochs in SAGE-Var were taken.  Interval 10 shows an unusually small number of variables because it is an order of magnitude shorter than any other interval, spanning only $\sim$20 days.  A randomly phased interval that short is unlikely to catch variability for many infrared sources.

\begin{figure}
\centerline{
\begin{tabular}{c}
\rotatebox{270}{\includegraphics*[width=0.65\linewidth]{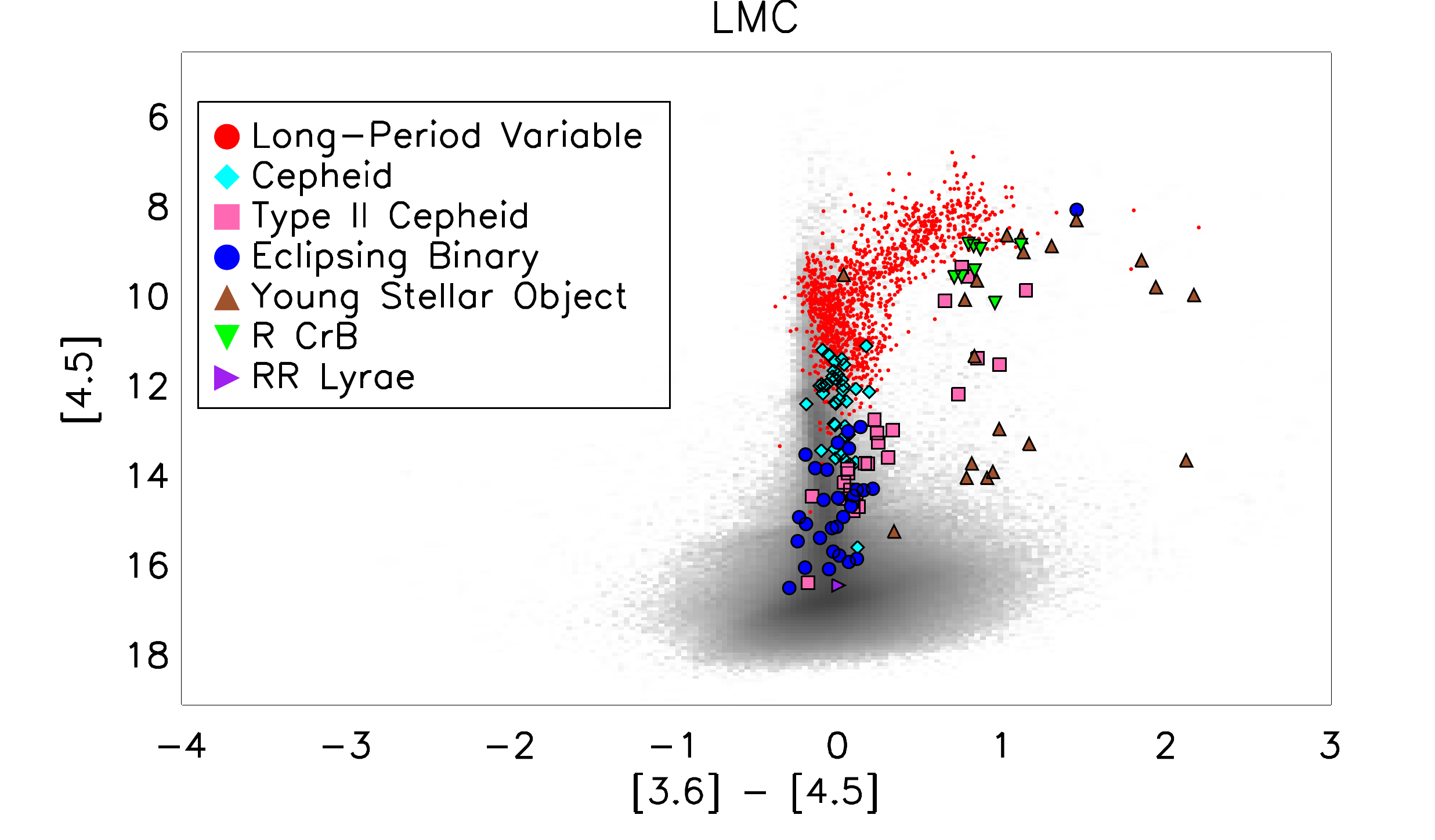}} \\
\rotatebox{270}{\includegraphics*[width=0.65\linewidth]{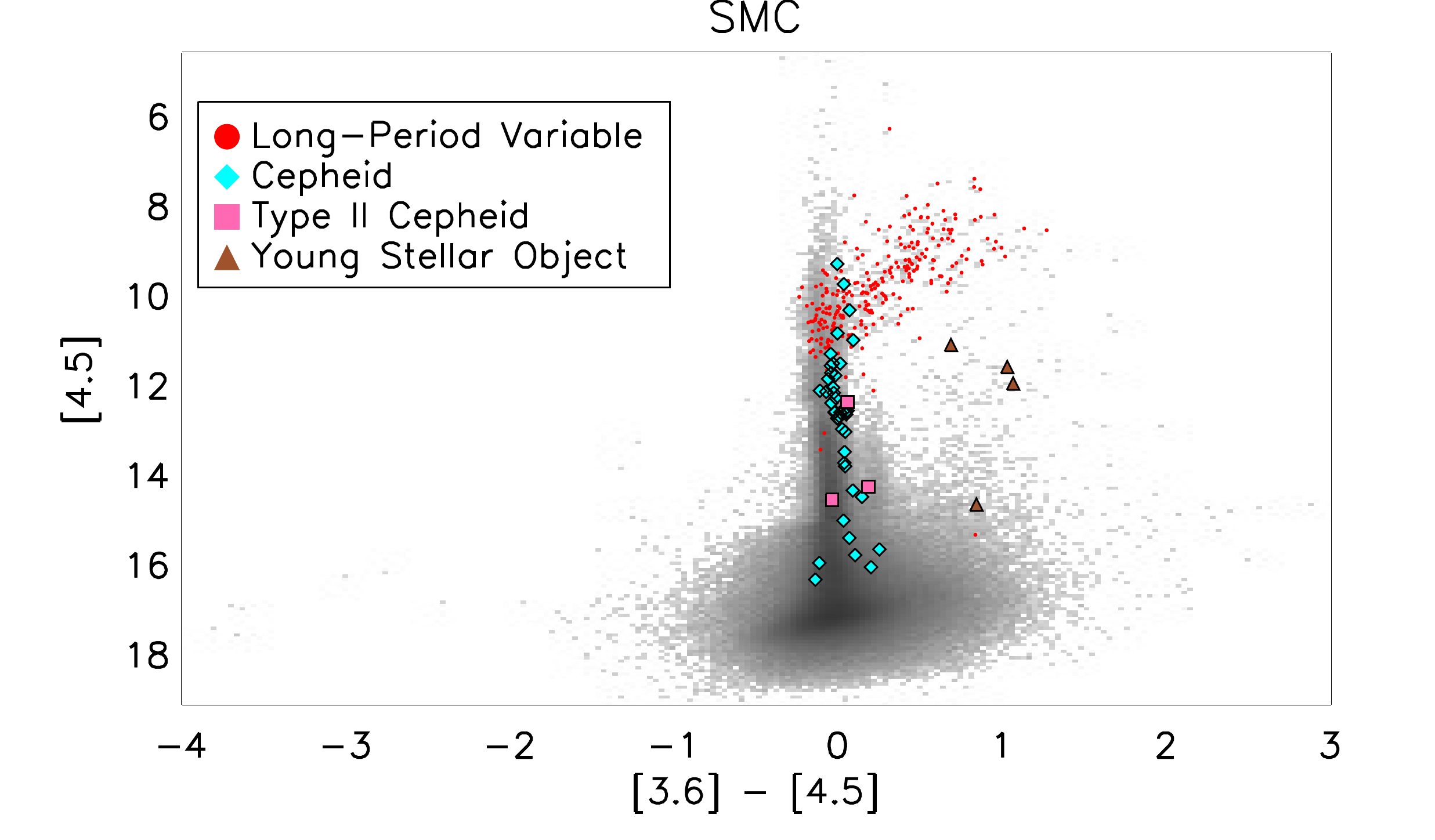}}
\end{tabular}
}
\caption[{[3.6] $-$ [4.5] vs.\ [4.5]} CMD of OGLE-CVS Variables]{[3.6] $-$ [4.5] vs.\ [4.5] CMD highlighting the OGLE-CVS variables with counterparts amongst the SAGE-Var classified variables.    The LMC sources are shown in the top panel and the SMC sources are shown in the bottom.  The entire SAGE-Var catalog is shown as a gray-scale Hess diagram in the background.  The same color scheme as in Figure~\ref{fig:var_hist} is used.}
\label{fig:var_cmd_ogle}
\end{figure}

In order to classify the variables we did detect, we matched our detected variables against the OGLE-III Catalog of Variable Stars (CVS)\footnote{ \url{http://ogledb.astrouw.edu.pl/~ogle/CVS/}}, which consists of $\sim$150,000 classified variables in the LMC ($\sim$26,500 in the SMC) with well characterized variability information.  Again using a 2\arcsec\ matching radius, we find 1361 OGLE-CVS matches to our \numlvars\ SAGE-Var variables in the LMC, and 323 matches to our \numsvars\ variables in the SMC.  These matches are shown in the [3.6] $-$ [4.5] vs.\ [4.5] CMD in Figure~\ref{fig:var_cmd_ogle}.  The OGLE populations we find in our data are detailed in Tables~\ref{tab:var_types_lmc} \&\ \ref{tab:var_types_smc}, along with the appropriate reference to the relevant OGLE-CVS document, if any.  See the cited references for thorough discussion and definitions of the various classifications.  The OGLE observations were taken primarily in the $I$ band, with some additional images in $V$ and $B$.  The OGLE camera has a pixel scale of $0.417\arcsec$/pixel, and concentrates 80\% of the light in a $0.5\arcsec$ disk.  See \citet{Udalski1997} for a thorough discussion of the OGLE hardware and observing protocols.

LPVs, mainly evolved AGB stars, are the most numerous OGLE sources we detect in our sample.  This is not surprising, as AGB stars are among the brightest objects in the IR sky, and many of our intervals probe timescales on which LPVs are expected to vary.  Sources identified as LPVs previously have been matched to variability data from the MACHO survey \citep{Riebel2010}.  The MACHO survey utilized two non-standard bands; a red ($\sim$690\,\mic) and a blue ($\sim$520\,\mic).  The MACHO optics have a pixel scale of $0.635\arcsec$/pixel and a median PSF of $2\arcsec$.  Details of the MACHO project and its photometry can be found in \citet{Alcock1997} and \citet{Alcock1999}.

We detect OGLE variables of every other OGLE classification besides LPVs as well, down to nearly the limiting magnitude of our survey.  These other classes tend to be blue in the diagrams of Figure~\ref{fig:var_cmd_ogle}, as the instability strip they occupy on the HR diagram places the peak of their SEDs blueward of the \textit{Spitzer} bands.

We also matched our data to the list of Young Stellar Object candidates (YSOs) in the LMC (SMC) compiled by \citet{Carlson2012} \citep{Sewilo2013}.  We find 500 (337) YSOs in the entire SAGE-Var LMC (SMC) dataset, but only 12 (4) of them are identified as variables using our criteria.  These are also shown in Figure~\ref{fig:var_cmd_ogle}.

\begin{figure}
\centerline{
\begin{tabular}{c}
\rotatebox{270}{\includegraphics*[width=0.65\linewidth]{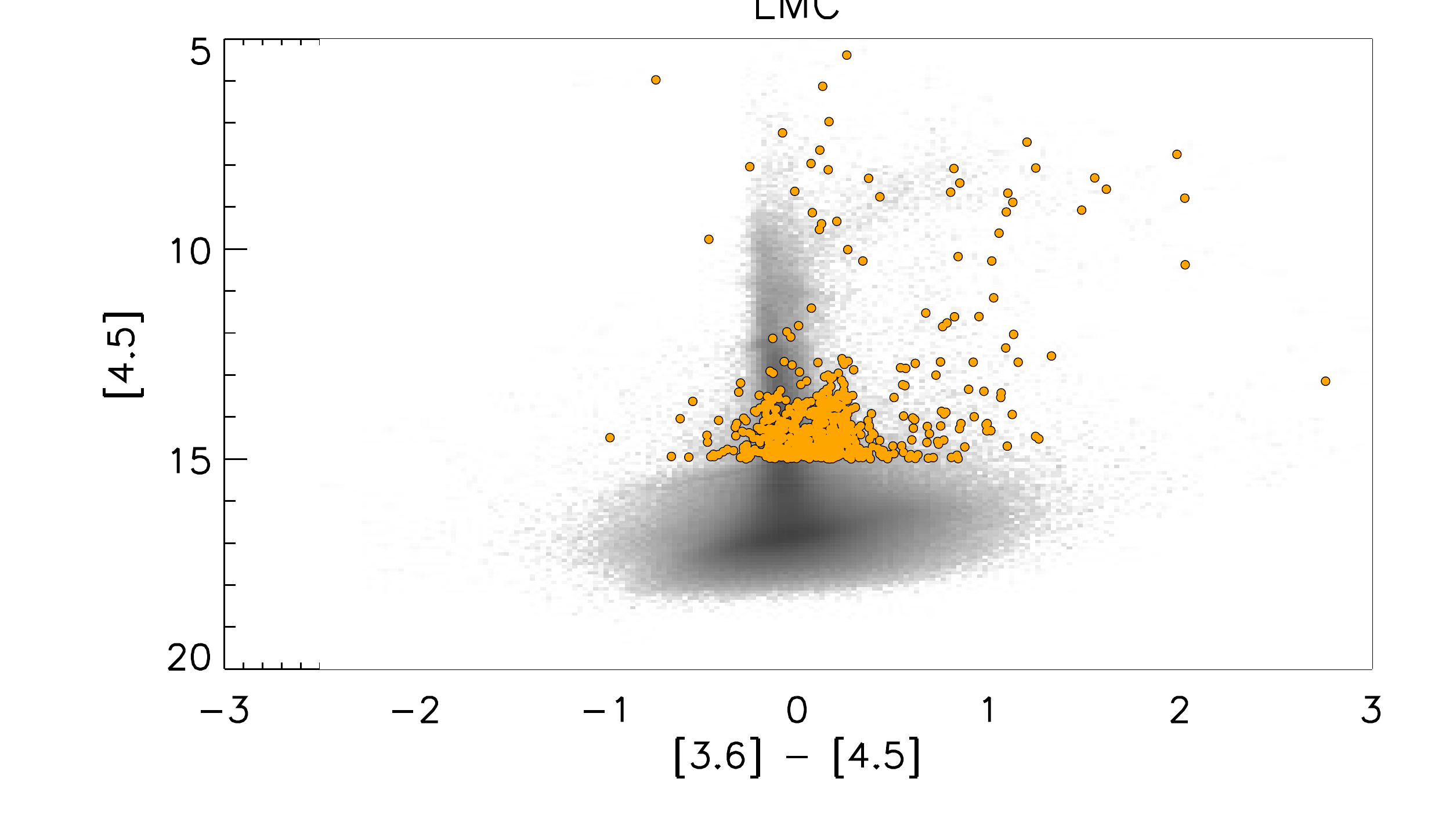}} \\
\rotatebox{270}{\includegraphics*[width=0.65\linewidth]{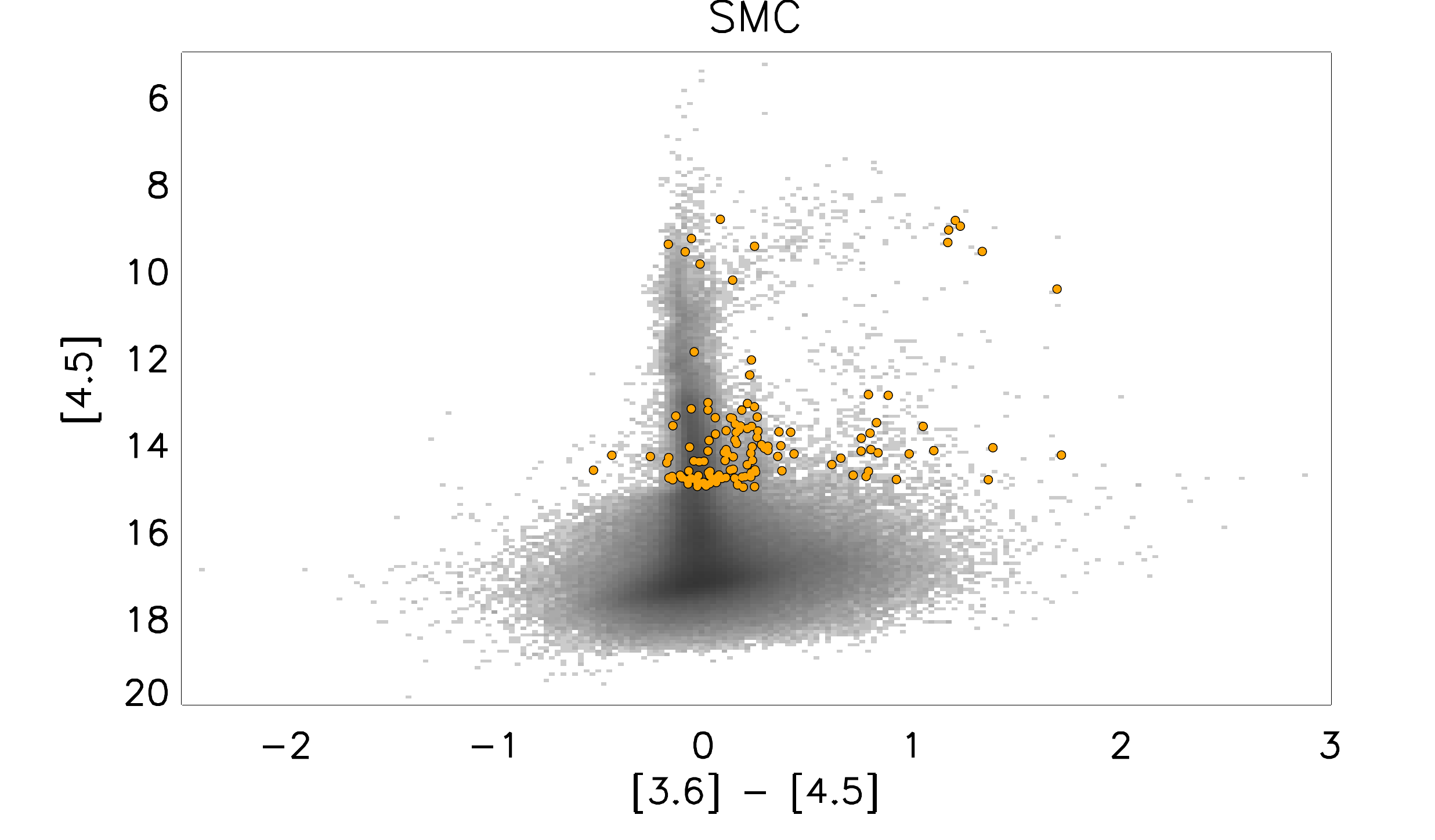}}
\end{tabular}
}
\caption[{[3.6] $-$ [4.5] vs.\ [4.5]} CMD of SAGE-Var variables with no OGLE counterpart]{[3.6] $-$ [4.5] vs.\ [4.5] CMD highlighting variables detected by SAGE-Var but not identified by the OGLE or MACHO surveys.  The LMC sources are shown in the top panel, and the SMC sources are shown in the bottom.  The entire SAGE-Var catalog is shown as a gray-scale Hess diagram in the background.  To eliminate false positives due to low S/N, we only consider sources brighter than 15$^{\rm th}$ magnitude to be variables.}
\label{fig:var_cmd_notogle}
\end{figure}

Figure~\ref{fig:var_cmd_notogle} shows the same CMDs as Figure~\ref{fig:var_cmd_ogle} but highlighting the sources without counterparts in the OGLE CVS or MACHO survey.  We remove any source with a SAGE [4.5] mosaic photometry dimmer than $15^{\mathrm{th}}$ magnitude identified as a variable but without an OGLE or MACHO counterpart.  We include these sources in the online catalog for completeness, but data artifacts (such as blending with nearby sources) cause problems with our variability criteria for such dim sources.  After removing those dim sources, we are left with \numunvarsl\ (\numunvarss) IR variables in the LMC (SMC) without OGLE or MACHO detected variation.  There are a few extremely red sources in the AGB region of the CMD without OGLE identifications.  Most of these sources are flagged as probable variables by the WISE survey, and our results bolster that conclusion (\S\,\ref{sec:new_lpvs}).  

\begin{deluxetable*}{lcl}
\tabletypesize{\scriptsize}
\tablecaption{Matched variable populations detected in SAGE-Var LMC}
\tablewidth{0pt}
\tablehead{\colhead{Variable Classification} & \colhead{Number} & \colhead{Reference/Definition}}
\startdata
Total Variables Detected & \numlvars \\
Long-Period Variables    & 1065 & \citet{ogle_lpv_lmc} \\
\hspace{0.5cm}AGB C & 5 \\
\hspace{0.5cm}AGB O & 6 \\
\hspace{0.5cm}Mira  AGB C & 426 \\
\hspace{0.5cm}Mira  AGB O & 143 \\
\hspace{0.5cm}OSARG AGB C & 15& OGLE small-amplitude variable reg giant (OSARG) \\
\hspace{0.5cm}OSARG AGB O & 42 \\
\hspace{0.5cm}OSARG RGB C & 1 \\
\hspace{0.5cm}OSARG RGB O & 17 \\
\hspace{0.5cm}SRV   AGB C & 262 & Semi-regular variable (SRV)\\
\hspace{0.5cm}SRV   AGB O & 148 \\
Cepheids                 &   28  & \citet{ogle_cep_lmc} \\
\hspace{0.5cm}1O &            1 & $1^{\mathrm{st}}$ overtone pulsation mode\\
\hspace{0.5cm}1O/2O &            1 & Mixture of $1^{\mathrm{st}}$ and $2^{\mathrm{nd}}$ overtone pulsation modes \\
\hspace{0.5cm}F &           26 &Fundamental-mode pulsation \\
Type II Cepheids         &   19 & \citet{ogle_t2_lmc} \\
\hspace{0.5cm}BLHer &            1 \\
\hspace{0.5cm}RVTau &           12 \\
\hspace{0.5cm}WVir &           6 \\
RR Lyrae                 &    3 & \citet{ogle_rrly_lmc} \\
R CrB                  &    6 & \citet{ogle_rcb_lmc} \\
Eclipsing Binaries       &   25 & \citet{Graczyk2011} \\
\hspace{0.5cm}EC &            8  & Contact binaries\\
\hspace{0.5cm}ECL &            1 & Ellipsoidal contact binaries\\ 
\hspace{0.5cm}ED &            8 & Detached binaries\\
\hspace{0.5cm}ESD &            8 & Semi-detached binaries\\
Young Stellar Objects    &   12 & \citet{Carlson2012}
\enddata
\tablecomments{This table lists the variable classifications and subclasses used by the OGLE Catalog of Variable Stars and the YSO catalog of \citet{Carlson2012}, and the numbers of such sources also detected as variable by the SAGE-Var survey in the LMC.  A 2\arcsec\ matching radius was used for all OGLE catalog comparisons, and the YSO list was matched based on SAGE Archive designations.  See the cited references for complete definitions of the subcategories listed here.  Uncommon abbreviations are defined in the third column.}
\label{tab:var_types_lmc}
\end{deluxetable*}

\begin{deluxetable}{lcr}
\tabletypesize{\small}
\tablecaption{Matched variable populations detected in SAGE-Var SMC}
\tablewidth{0pt}
\tablehead{\colhead{Variable Classification} & \colhead{Number} & \colhead{Reference}}
\startdata
Total Variables Detected & 571 \\
Long-Period Variables    &  276 & \citet{ogle_lpv_smc} \\
\hspace{0.5cm}Mira  C &          140 \\
\hspace{0.5cm}Mira  O &           22 \\
\hspace{0.5cm}OSARG C &            7 \\
\hspace{0.5cm}OSARG O &            4 \\
\hspace{0.5cm}SRV   C &           95 \\
\hspace{0.5cm}SRV   O &            8 \\
Cepheids                 &   42 & \citet{ogle_cep_smc} \\
\hspace{0.5cm}F &           40 \\
\hspace{0.5cm}1O &            2 \\
Type II Cepheids         &    3 & \citet{ogle_t2_smc} \\
\hspace{0.5cm}RVTau &            1 \\
\hspace{0.5cm}WVir &            1 \\
\hspace{0.5cm}pWVir &            1 \\
Young Stellar Objects & 4 & \citet{Sewilo2013} 
\enddata
\tablecomments{This table lists the variable classifications and subclassifications used by the OGLE Catalog of Variable Stars and the YSO catalog of \citet{Sewilo2013}, and the numbers of such sources also detected as variable by the SAGE-Var survey in the SMC.  A 2\arcsec\ matching radius was used for all OGLE catalog comparisons, and the YSO list  was matched based on SAGE Archive designation.}
\label{tab:var_types_smc}
\end{deluxetable}

\section{RESULTS} \label{sec:results}

\subsection{New AGBs}\label{sec:new_lpvs}
\begin{figure}
\centerline{
\begin{tabular}{c}
\rotatebox{270}{\includegraphics*[scale=0.40]{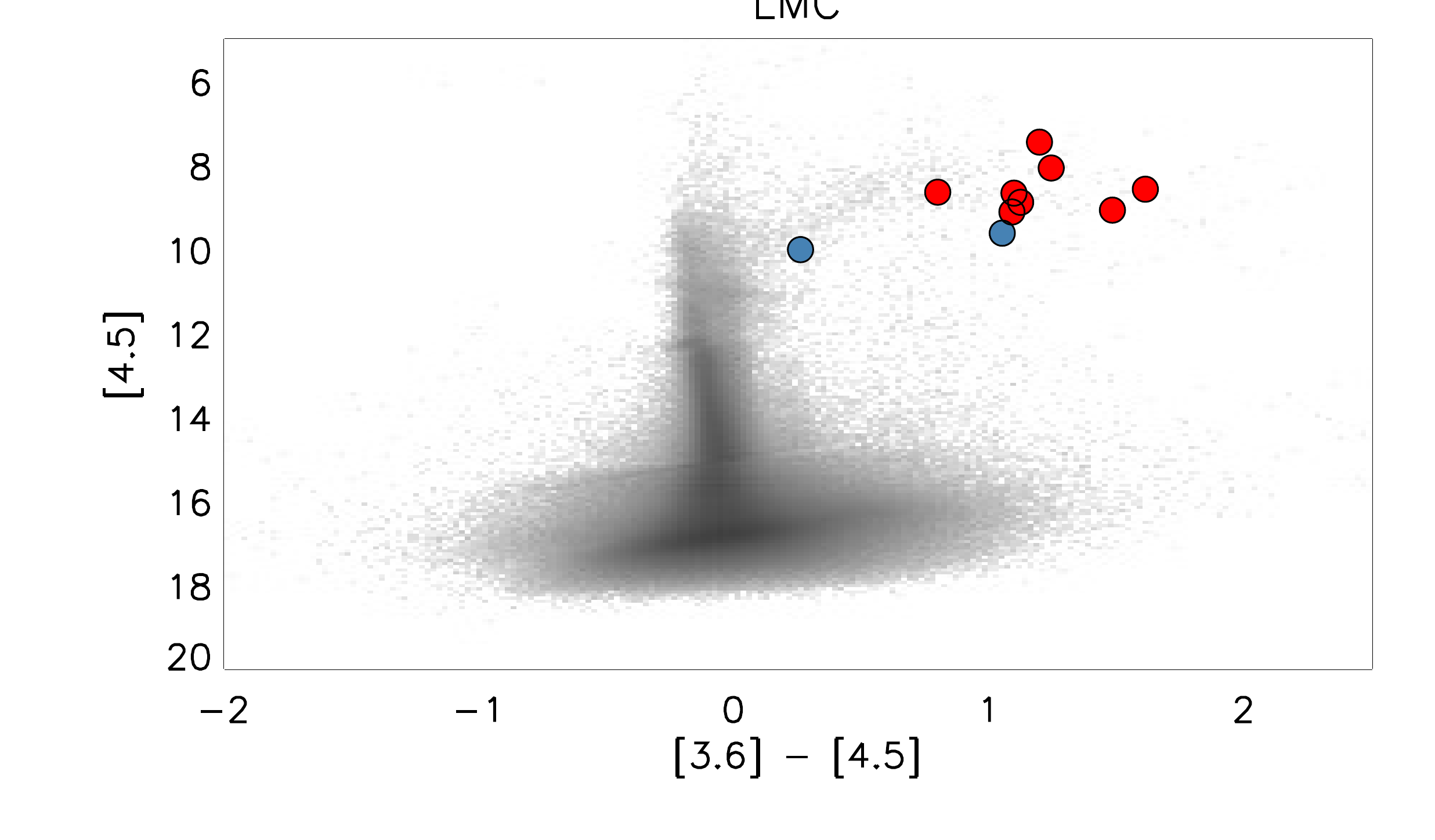}} \\
\rotatebox{270}{\includegraphics*[scale=0.40]{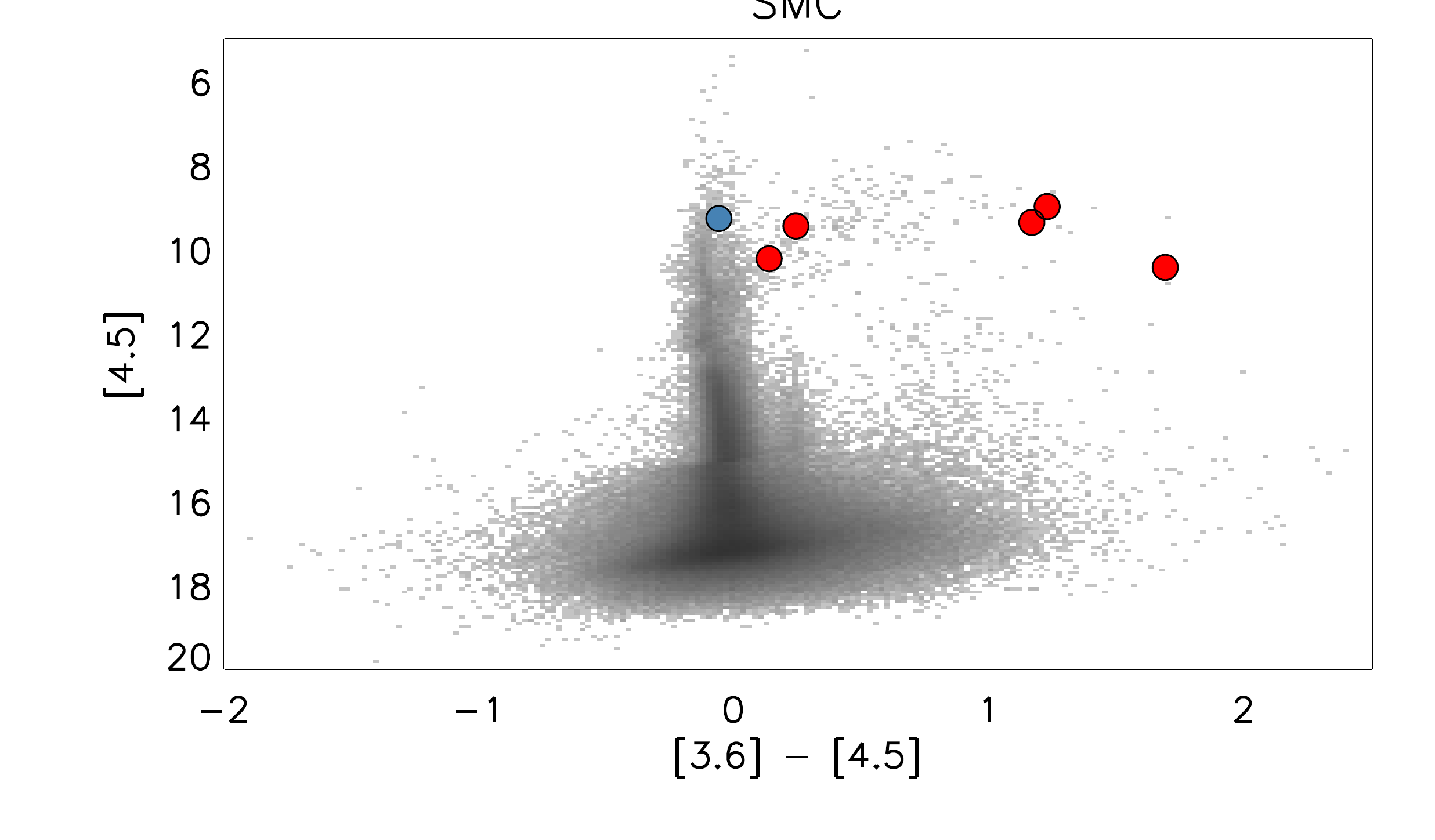}}
\end{tabular}
}
\caption{$[3.6] - [4.5]$ vs.\ [4.5] CMD highlighting newly identified LPV candidates.  The LMC sources are shown in the top panel, and the SMC sources are shown in the bottom.  The entire SAGE-Var catalog is shown as a gray-scale Hess diagram in the background.  Stars classified as C-rich by the GRAMS model grid are shown in red, stars classified as O-rich are shown in blue.}
\label{fig:new_lpv_cmd}
\end{figure}

Because they typically require large amounts of telescope time and the expense of space-based multi-epoch observations often makes such lengthy surveys cost-prohibitive, most variability surveys are conducted from the ground, and hence at optical wavelengths.  Ground-based variability surveys thus miss the reddest, most extreme AGBs, exactly the stars whose variability is most relevant to the evolved star dust budget because they dominate the mass return from AGBs to the ISM \citep{Riebel2012, Boyer2012, Matsuura2009}.  SAGE-Var represents the first large-scale variability survey at such red wavelengths.  As such, we have the ability to detect variable stars which have never been categorized as such before.  Our survey detects variability in  \numunvarsl\ (\numunvarss) sources in the LMC (SMC) which do not have well-defined variability measurements in the OGLE or MACHO surveys (Fig.~\ref{fig:var_cmd_notogle}).  Out of these newly identified variables, 10 sources  in the LMC and 6 in the SMC can be classified as AGB candidates, based on previous studies \citep{Riebel2012}, or their position in the \jmag\ $-$ \ks\ vs. \ks\ CMD.  Figure~\ref{fig:new_lpv_cmd} highlights these 10 (6) new AGB candidates against a Hess diagram of the entire SAGE-Var sample. 

\subsubsection{LMC AGB Candidates} \label{subsec:lmc_lpvs}
We find 10 AGB candidates in the LMC without previously well-measured periods in either the OGLE or MACHO surveys.  Most of these were flagged as highly likely to be true variables by the WISE survey, and we confirm that measurement.  We do not find any of these 10 candidates in the SAGE-Spec \citep{Kemper2010} list of LMC sources.  Without a chemical classification derived from observed spectra, we therefore employ an Spectral Energy Density (SED)-based chemical classification.

We fit the SEDs of these sources with synthetic SEDs from the GRAMS model grid \citep{Sargent2011, Srinivasan2011}.  GRAMS is a grid of dusty AGB/RSG star models computed by the radiative transfer code 2Dust \citep{Ueta2003}.  The grid is constructed by placing spherically symmetric circumstellar dust shells of varying inner radii, optical depth, and both O-rich and C-rich chemical composition around model stellar photospheres.  The models assume a constant mass-loss rate ({\it i.e.}, an inverse-square density law) in the shell and a power-law distribution of grain sizes with an exponential falloff, with a typical size of $\sim$0.1\,$\mu$m.  The O-rich dust is modeled using the astronomical silicates of \citet{Ossenkopf1992}, and the carbonaceous dust is a 9:1 mixture by mass of amorphous carbon \citep[optical constants from][]{Zubko1996} and silicon carbide \citep[optical constants from][]{Pegourie1988}.  We perform minimum $\chi^2$ fitting to obtain the best-fit parameters, which include the luminosity, dust-production rate and chemical type.

 Using the GRAMS model grid, we classify 8/10 of these AGB candidates as C-rich.  C-rich AGB stars tend to be redder than O-rich stars, and our newly measured variables skew red.  AGB stars in our sample detected by the OGLE survey have an average [3.6] $-$ [4.5] color of $-0.07$\,mag, while the sources with no OGLE detection are nearly a magnitude redder at these wavelengths, with an average [3.6] $-$ [4.5] of 0.74\,mag.   SAGE-Var has too sparse a sampling of the lightcurves to deduce a period for their variability, but we can place lower limits on their IR variability amplitudes.  These sources are listed in Table~\ref{tab:new_lpvs_lmc}.

\begin{deluxetable*}{lcccccc}
\tabletypesize{\small}
\tablecaption{AGB Candidates in the LMC without OGLE or MACHO Variability Measurement}
\tablewidth{0pt}
\tablehead{\colhead{SAGE} & \colhead{RA} & \colhead{Dec}  & \colhead{GRAMS} & \colhead{[3.6]} & \colhead{[4.5]}&\colhead{[3.6]--[4.5]} \\ \colhead{Designation\tablenotemark{a}} & \colhead{(2000)} &\colhead{(2000)} & \colhead{Class} & \colhead {Amplitude} & \colhead{Amplitude} & \colhead{color}}
\startdata
J051041.21-683606.6 &       77.6717 &      -68.6018 & C & 
     0.83 &      0.85 &       1.24 \\
J051414.85-700409.8 &       78.5619 &      -70.0694 & C & 
     0.93 &      0.86 &       1.12 \\
 J052503.26-692617.3 &       81.2636 &      -69.4381 & C & 
     0.90 &      0.69 &       1.48 \\
J052813.02-691228.4 &       82.0543 &      -69.2079 & C & 
     0.86 &      0.75 &       1.09 \\
 J052900.19-695247.3 &       82.2508 &      -69.8798 & C & 
      1.11 &      0.52 &      0.80 \\
  J053051.75-694328.0 &       82.7156 &      -69.7245 & C & 
      1.42 &       1.19 &       1.20 \\
J050202.38-690726.2 &       75.5099 &      -69.1239 & C & 
     0.52 &      0.56 &       1.10 \\
 J050718.89-683850.4 &       76.8287 &      -68.6474 & C & 
      1.50 &       1.27 &       1.61 \\
J051913.89-693818.3 &       79.8079 &      -69.6384 & O & 
     0.29&      0.31 &      0.26 \\
J053010.30-690933.8 &       82.5429 &      -69.1594 & O & 
     0.36 &      0.30 &       1.05 
\enddata
\tablenotetext{a}{Designations in the online data table are prefaced with `SSTISAGEMA'}
\tablecomments{The 10 AGB candidates in the LMC with variability newly detected by SAGE-Var.  Many of these sources are confirmed to be AGB stars in the literature, but none have previously been observed to vary.  The GRAMS Class column lists the classification (O-rich or C-rich) assigned each source by the GRAMS model grid.  The next two columns represent a lower bound on the IR variability amplitude of the sources, the maximum observed magnitude minus the minimum observed magnitude.  The $[3.6] - [4.5]$ color is also listed to connect the entries in this table to Figure~\ref{fig:new_lpv_cmd}.}
\label{tab:new_lpvs_lmc}
\end{deluxetable*}

\subsubsection{SMC AGB Candidates}
After cross-matching our list of AGB candidates in the SMC with SIMBAD and removing all sources confirmed to not be AGB stars, we generate a list of 6 new AGB candidates.  We verified that none of these candidates were in the Ruffle et al. (2015, \emph{in prep.}) list of spectroscopically classified SMC sources.  Using the GRAMS model grid, we classify 5/6 of them as C-rich.  These sources are listed in Table~\ref{tab:new_lpvs_smc}.  Source SSTISAGEMA J010041.61-723800.7 is the reddest new LPV candidate we idenitfy in the SMC.  The best-fitting GRAMS model is an O-rich model, but the extreme redness of the source is more consistent with a C-rich star.  Currently, the GRAMS model grid in the SMC places too much weight on the 24\,\mic\ photometry, and this leads to some sources being mis-classified (see Srinivasan, et al. 2015, in prep for details).  We manually change the classification of this source to C-rich.

\begin{deluxetable*}{lcccccc}
\tabletypesize{\small}
\tablecaption{AGB Candidates in the SMC without OGLE or MACHO Variability Measurement}
\tablewidth{0pt}
\tablehead{\colhead{SAGE} & \colhead{RA} & \colhead{Dec}  & \colhead{GRAMS} & \colhead{[3.6]} & \colhead{[4.5]}&\colhead{[3.6]--[4.5]} \\ \colhead{Designation\tablenotemark{a}} & \colhead{(2000)} &\colhead{(2000)} & \colhead{Class} & \colhead {Amplitude} & \colhead{Amplitude} & \colhead{color}}
\startdata
 J005106.28-731635.9 &       12.7762 &      -73.2767 & C & 
     0.34 &      0.34 &      0.14 \\
J004544.12-720815.4 &       11.4338 &      -72.1376 & C & 
     0.32 &      0.40 &      0.24 \\
 J005926.35-722341.4 &       14.8598 &      -72.3949 & C & 
      1.09 &      0.88 &       1.23 \\
J010232.75-721912.5 &       15.6365 &      -72.3202 & C & 
      1.08 &       1.06 &       1.17 \\
J010041.61-723800.7 &       15.1734 &      -72.6335 & C\tablenotemark{b} & 
     0.95 &       1.03 &       1.69 \\
J005131.21-732007.7 &       12.8801 &      -73.3355 & O & 
     0.34 &      0.23 &    -0.06
\enddata
\tablenotetext{a}{Designations in the online data table are prefaced with `SSTISAGEMA'}
\tablenotetext{b}{The best-fitting GRAMS model for this source is an O-rich model with poor fit quality.  Based on the extreme redness of this source, we classify it as C-rich}
\tablecomments{The 6 AGB candidates in the SMC with variability newly detected by SAGE-Var.  Most of these sources are confirmed AGB stars, but none have previously been observed to vary.  The GRAMS Class column lists the classification (O-rich or C-rich) assigned each source by the GRAMS model grid (Srinivasan, et al., 2015, in prep).  The next two columns represent a lower bound on the IR variability amplitude of the sources, the maximum observed magnitude minus the minimum observed amplitude.  The $[3.6] - [4.5]$ color is also listed to connect the entries in this table to Figure~\ref{fig:new_lpv_cmd}.}
\label{tab:new_lpvs_smc}
\end{deluxetable*}

\subsection{Variability Amplitude and Dust Production Rate}\label{sec:amp_dpr}
\citet{Le Bertre1992} found a strong correlation between variability amplitude (measured in the $K$ band), and $K-L^{\prime}$ color for a sample of 20 carbon-rich LPVs in our galaxy.  \citet{Whitelock1991} and \citet{Le Bertre1993} found a similar correlation for O-rich LPVs.  \citet{Whitelock1994} found a strong correlation between the IR amplitude and the $K -  [12]$ color.  The $K$ band is dominated by light from the stellar photosphere, while the IRAS 12\,\mic\ band represents the emission from the cooler dust shell around the star, and so this color serves as an indicator of the thickness of the circumstellar dust shell, and hence the rate of dust production by the star, if dust production rate (DPR) is assumed to be constant.

\begin{figure} 
\centerline{
\rotatebox{270}{\includegraphics*[width=0.65\linewidth]{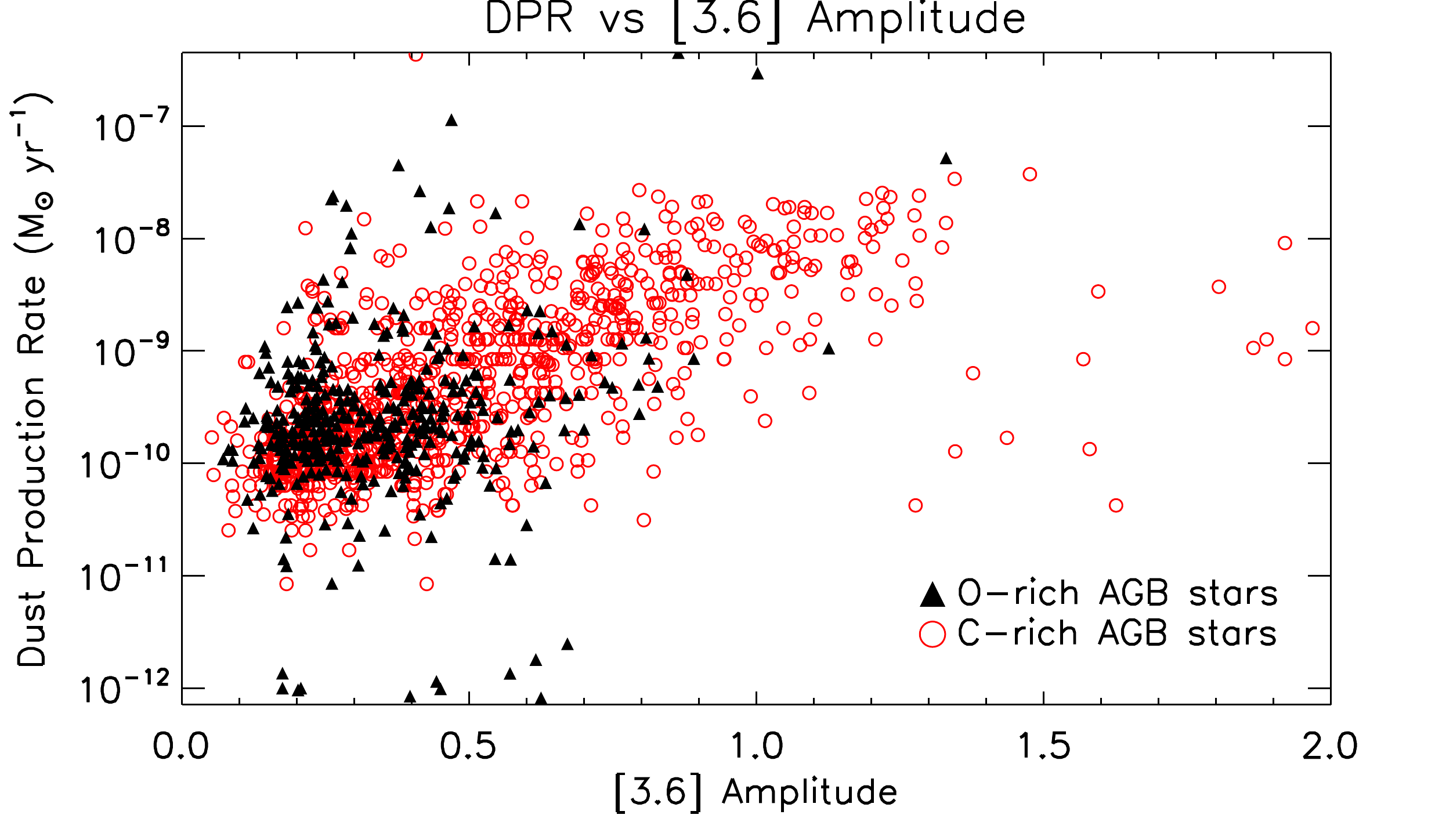}}
}
\caption{Dust Production Rate (DPR, \msunperyr) vs [3.6] Amplitude for AGB stars in the LMC and SMC.  The DPR is taken from the best-fitting GRAMS model, and the [3.6] Amplitude is defined as the dimmest measured SAGE-Var [3.6] magnitude minus the brightest [3.6] magnitude.  C-rich AGB stars are shown as open red circles, and O-rich AGB stars as filled black triangles.} \label{fig:amp_mlr}
\end{figure}

By using the GRAMS modeling results of \citet{Riebel2012}, we can directly compare the [3.6] amplitude and the DPR for our sample.  Our results are shown in Figure~\ref{fig:amp_mlr}.  We extract every source in the SAGE-Var catalog with a valid GRAMS classification which was also classified as a true variable (\S\,\ref{sec:source_class}).  We see a slight correlation of increasing DPR with increasing infrared amplitude (albeit one with a very large scatter).  C-rich and O-rich AGB stars have considerable overlap in their range of DPR, but more C-rich stars than O-rich stars extend to higher DPRs \citep[see discussion in][]{Riebel2012}.  Because of this, the trend is more visible among the C-rich sources, but both populations follow the same basic trend, with considerably more scatter at low amplitudes for the O-rich sources.

\subsection{Period-Luminosity Relationships}\label{sec:pl}
\subsubsection{AGB Period-Luminosity Relationships}\label{subsec:agb_pl}
Characterizing the AGB Period-Luminosity (PL) relationship is important in light of the upcoming launch of the James Webb Space Telescope (JWST).  AGB stars are among the brightest sources in the infrared (IR) sky, and with a large orbital platform concentrating on the IR, AGB stars could serve as important distance indicators, if their intrinsic luminosity can be well determined from their variability.  Previous investigations of the AGB PL relationship \citep[see][and references therein]{Riebel2010} have found a scatter of a few tenths of a magnitude about the best-fit line, and it has not been clear whether this scatter was intrinsic to the relationship (i.e. a real phenomenon due to an astrophysical cause such as a spread in stellar masses) or an artifact resulting from observing stars at various points in their light curves.  Constructing mean magnitudes from observations taken at several points in an AGB star's lightcurve are an essential step toward clarifying the nature of this spread.  The 6 epochs of the SAGE-Var survey allow us to calculate improved mean magnitudes for every source in the catalog.

We use only AGB stars from the sample of \citet{Riebel2010} which are identified as variables in the SAGE-Var data to construct our relations.  Only stars that have been observed at more than one brightness benefit from the averaging process, as opposed to stars with 6 measurements of the same brightness.  For all AGB stars identified as variables, all available valid flux measurements from the SAGE and SAGE-Var surveys were averaged to construct mean fluxes at both [3.6] and [4.5].  Photometric errors from each observation were added in quadrature to produce an error for the mean flux.  These mean fluxes were transformed into magnitudes using the zero points given in \S\,\ref{sec:reduction}.

\begin{figure}
\centerline{
\begin{tabular}{c}
\rotatebox{270}{\includegraphics*[scale=0.40]{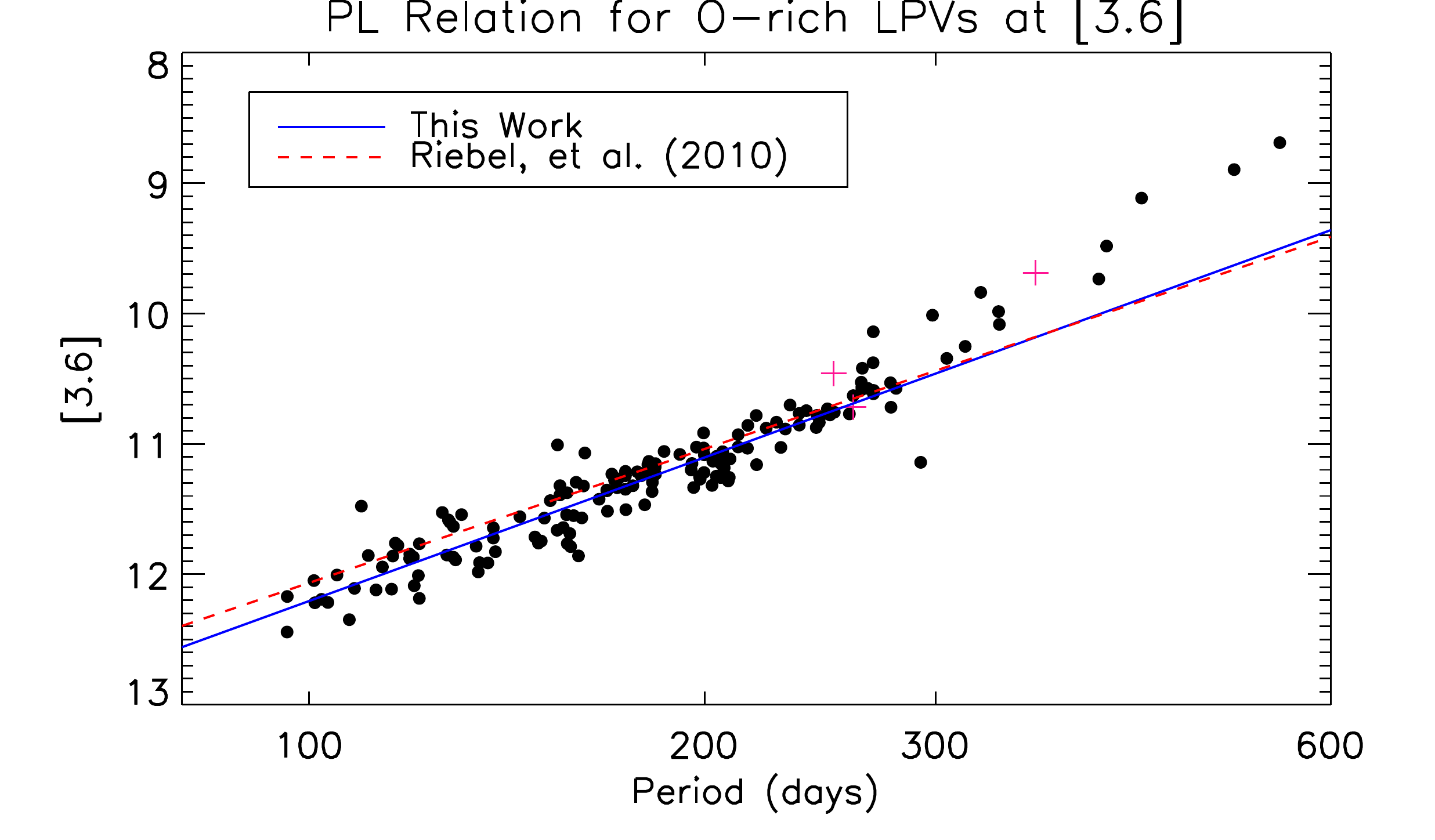}} \\
\rotatebox{270}{\includegraphics*[scale=0.40]{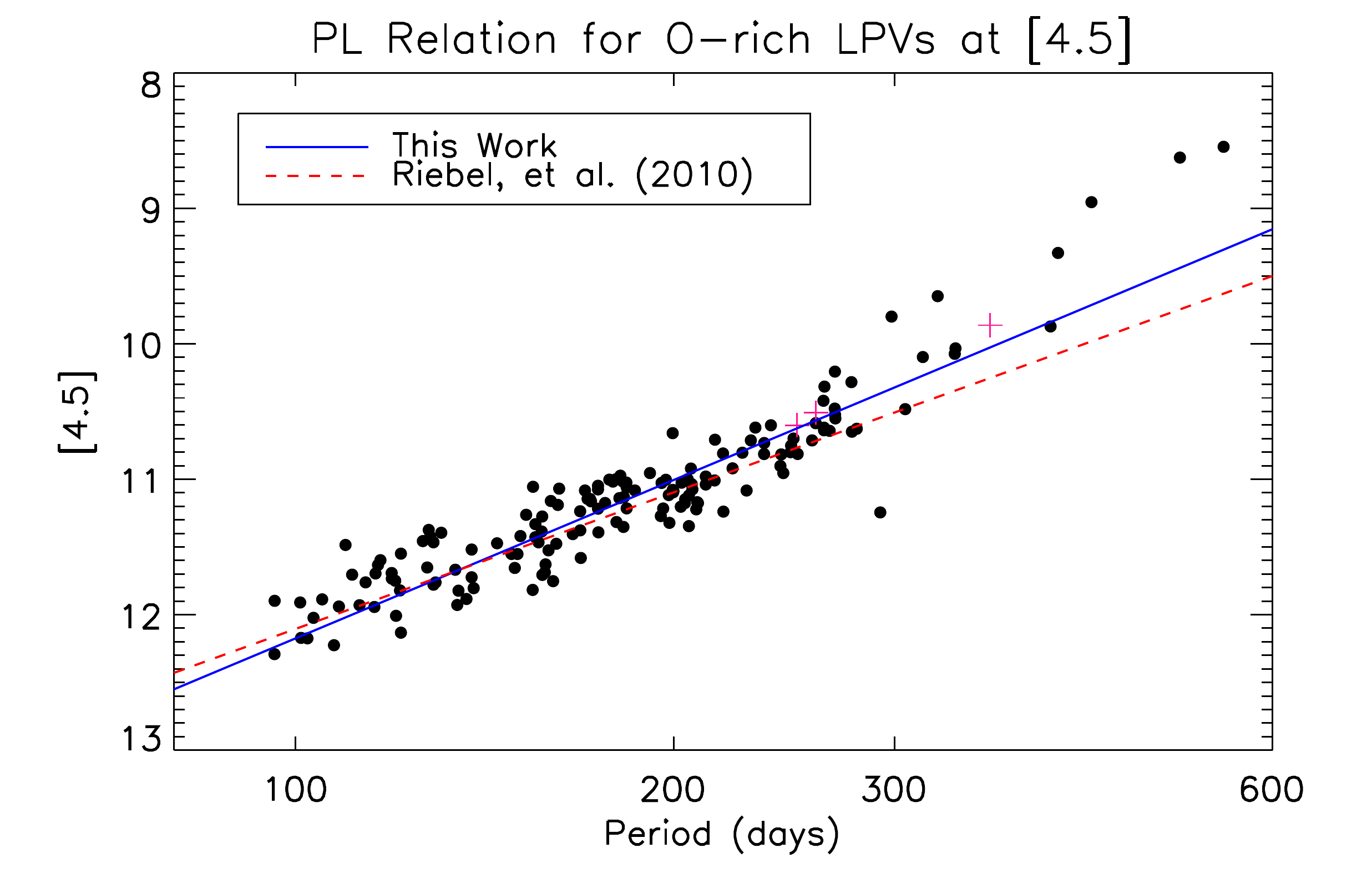}}
\end{tabular}
}
\caption[PL Relationships for AGB Stars in the LMC from SAGE-Var Mean Magnitudes]{PL relationships for GRAMS classified O-rich AGB stars in the SAGE-Var sample, constructed using 6-epoch mean magnitudes in the [3.6] and [4.5] bands.  Stars from the LMC are shown as black points, stars from the SMC shown as pink plus signs.  Note the horizontal axis is a log scale.}
\label{fig:oagb_pl}
\end{figure}

\begin{figure}
\centerline{
\begin{tabular}{c}
\rotatebox{270}{\includegraphics*[width=0.65\linewidth]{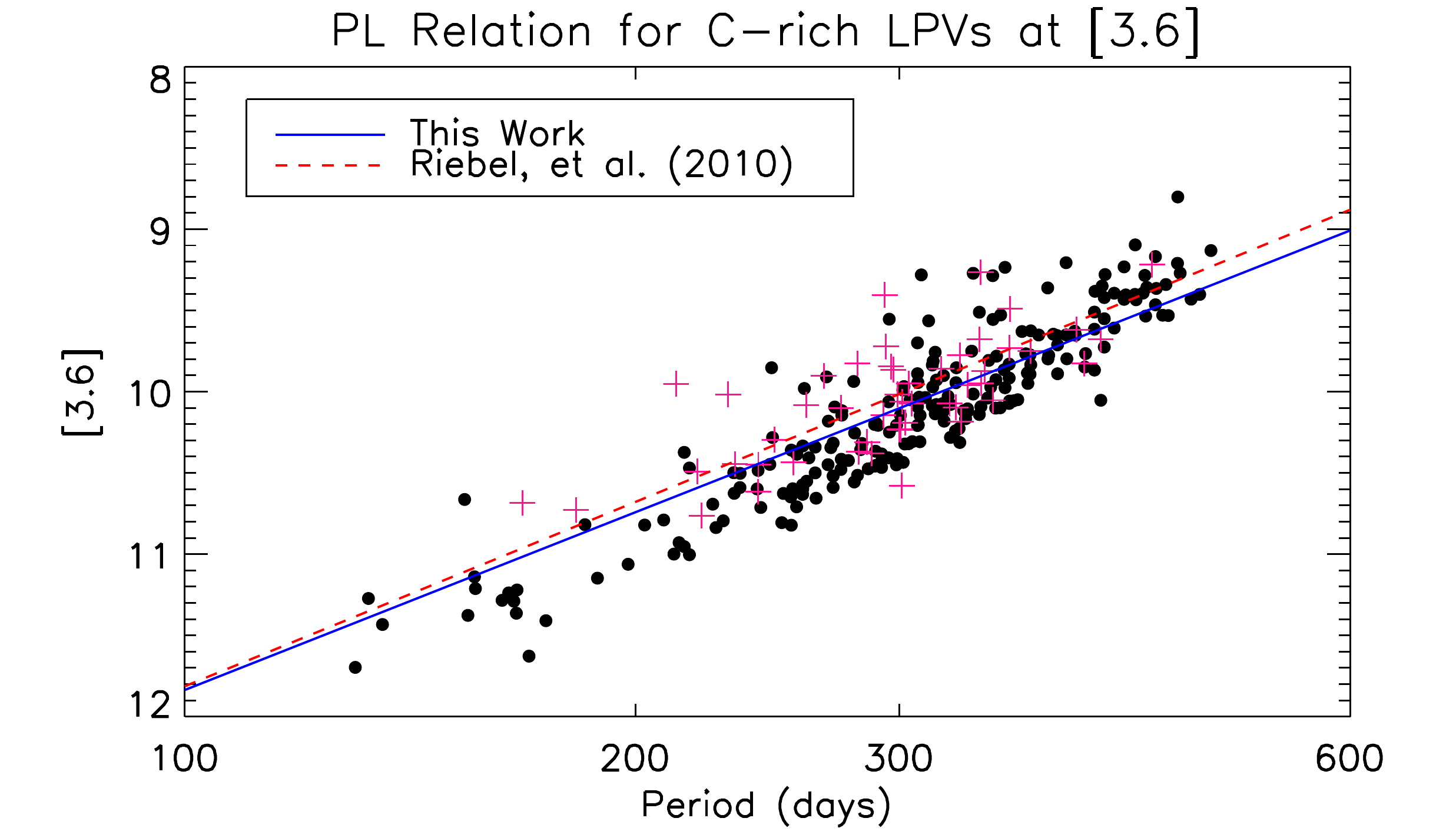}} \\
\rotatebox{270}{\includegraphics*[width=0.65\linewidth]{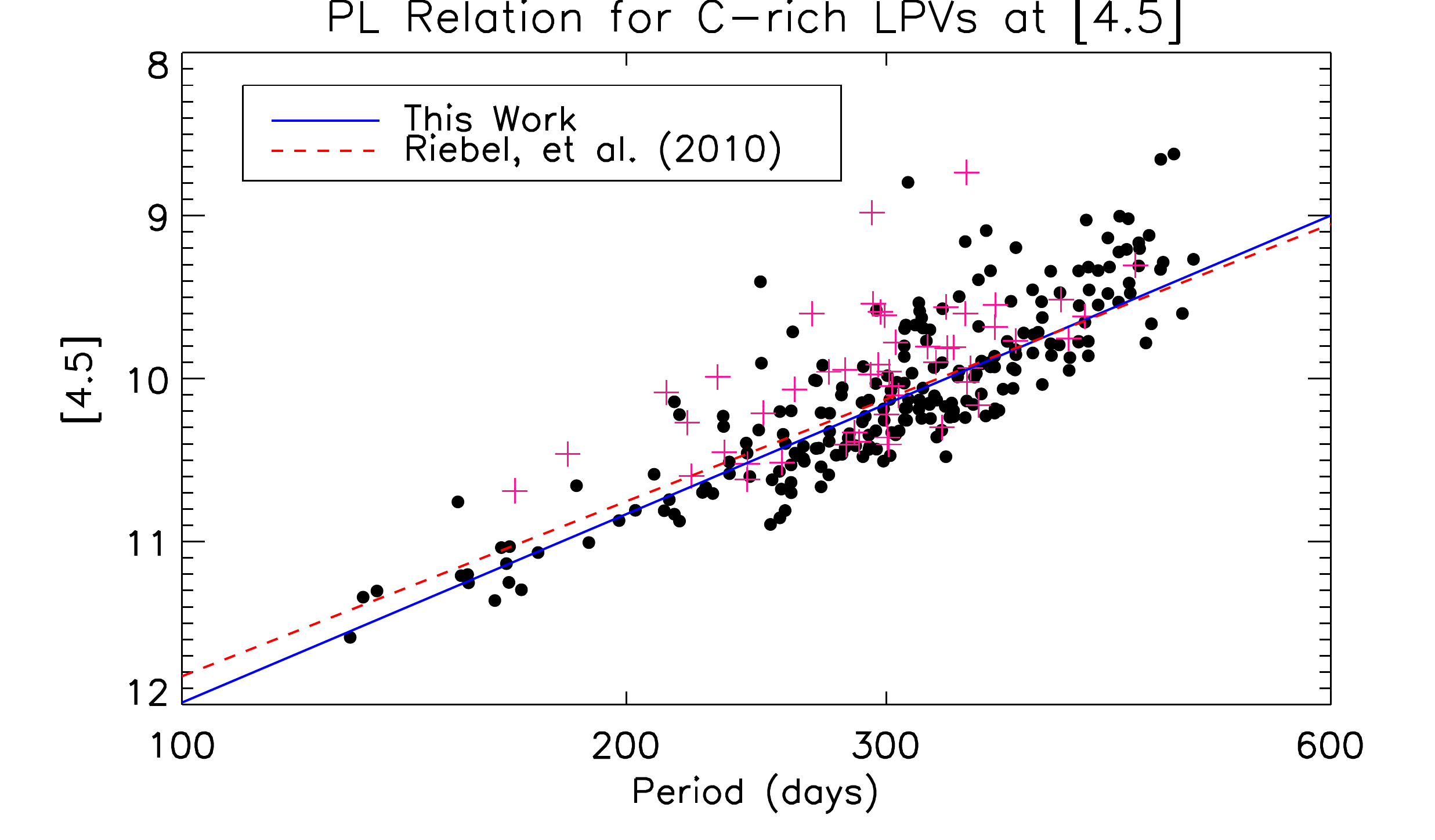}}
\end{tabular}
}
\caption[PL Relationship for AGB Stars in the SMC from SAGE-Var Mean Magnitudes]{PL relationships for GRAMS classified C-rich AGB stars in the SAGE-Var sample, constructed using 6-epoch mean magnitudes in the [3.6] band.  Stars from the LMC are shown as black points, stars from the SMC shown as pink plus signs.  Note the horizontal axis is a log scale.}
\label{fig:cagb_pl}
\end{figure}

Figure~\ref{fig:oagb_pl} shows the PL relations constructed from the 6-epoch mean magnitudes constructed from the SAGE-Var data.  The 3.6\,\mic\ magnitude is used in the top panel, and the 4.5\,\mic\ band is used in the bottom.  The periods come from the MACHO survey if possible, and from the OGLE survey if no MACHO period was available.  The MACHO periods were given priority in order to maintain consistency with the study of \citet{Riebel2010}.  For those stars with measured periods in both the OGLE and MACHO surveys, the periods were found to differ by only 3 days on average.  We only present updated fits for stars pulsating in the fundamental mode (sequence 1), as this was the only sequence with a significant population observed in SAGE-Var.  The derived PL relations for the stars classified as O-rich (by the best-fitting GRAMS model) are shown in Figure~\ref{fig:oagb_pl} and quantitatively described in Table~\ref{tab:oagb_pl}.  The fits to the stars identified as C-rich are shown in Figure~\ref{fig:cagb_pl} and described in Table~\ref{tab:cagb_pl}.

In Figure~\ref{fig:oagb_pl} stars with periods $\gtrsim300$\,days are systematically brighter than the trend line, whereas in Figure~\ref{fig:cagb_pl} stars with periods $\lesssim300$\,days are systematically dimmer than the trend line.  These outlier stars are an overlap between the two populations.  The C-rich best-fit trend line is about a magnitude brighter than the O-rich fit, and C-rich stars have longer periods on average.  Thus, the shortest period C-rich stars that are dimmer than the trend line actually overlap the main body of the O-rich stars in PL space.  In the same manner, the longest period O-rich stars which are brighter than the trend line in figure~\ref{fig:oagb_pl} overlap the main population of C-rich stars in figure~\ref{fig:cagb_pl}.  When using photometric classification, the division between O- and C-rich AGB stars is not sharp, and the two populations smoothly blend into one another in PL space \citep[see Figures 1 \& 5 in][]{Riebel2010}.  Because there are only about 10 of these outlying stars, the fits are insensitive to their inclusion, and they have not been removed from the sample prior to fitting.

Using the criteria described in Appendix A of \citet{Riebel2010}, we identify three O-rich AGB stars in the SMC which lie on Sequence 1.  We corrected for the difference in distance between the LMC and the SMC using distance moduli of 18.54 and 18.93, respectively \citep{Keller2006}.  These stars, shifted to the distance of the LMC, are plotted as pink crosses in Figure~\ref{fig:oagb_pl}.  On average, the LMC has a greater metallicity than that of the SMC.  Our numbers are too small to draw definite conclusions, but we do not see any evidence for a dependence of the AGB PL relation on metallicity in our sample.  This idea is worthy of further investigation.

While we do see a slight decrease in the scatter about the best-fit line compared to that found by \citet{Riebel2010}, the reduction is less than a factor of two, which leads us to believe much of this remaining scatter is intrinsic to the relationship and is not a product of observing different stars at different phases of their lightcurve.

\begin{deluxetable}{lccc}
\tabletypesize{\small}
\tablecaption{Period Magnitude Relationships for O-rich AGB stars Pulsating in the Fundamental Mode}
\tablewidth{0pt}
\tablehead{\colhead{PL Relation} & \colhead{N} & \colhead{scatter (mag)}}
\startdata
This work: \\
$[3.6] = -4.00(\pm 0.03) \log P + 20.24(\pm 0.07)$ & 131 & $0.18$ \\
$[4.5] = -3.78(\pm 0.03) \log P + 19.72(\pm 0.06)$ & 131 & $0.22$ \\
\citet{Riebel2010}: \\
$[3.6]=-3.41(\pm 0.04) \log P + 18.88(\pm 0.09)$ & 2221 & $0.271$ \\
$[4.5]=-3.35(\pm 0.04) \log P + 18.80(\pm 0.01)$ & 2227 & $0.270$
\enddata
\tablecomments{PL relations for LPVs classified as O-rich AGB stars.  The quoted scatter is the standard deviation of the residuals about the best-fit line.}
\label{tab:oagb_pl}
\end{deluxetable}

We identified 223 (49) Carbon-rich, non-extreme AGB stars pulsating in the fundamental mode in the LMC (SMC).  ``Extreme" AGB stars are defined as those stars with $J - [3.6] > 3.1$.  The flux in the \jmag\ band traces the emission from the stellar photosphere, while the [3.6] emission is from the circumstellar dust shell around these heavily enshrouded stars.  At SAGE-Var wavelengths, these stars do not follow the same PL relation as less enshrouded stars because their brightness does not reflect stellar emission, but emission from the dust shell \citep[see figure 4 in][]{Riebel2010}.  Sources thus classified as extreme are not included in this fit.  The derived PL relations to the remaining C-rich stars are shown in Figure~\ref{fig:cagb_pl} and quantitatively described in Table~\ref{tab:cagb_pl}.  The [4.5] data are shown for completeness, but we do not attribute any significance to the linear fits in that band.  As discussed by \citet{Riebel2010}, AGB stars are affected by a CO absorption feature in the [4.5] band which distorts the linear PL relationship in that band.  O-rich stars do not have enough CO in their atmospheres to be greatly affected. Figure 6 in \citet{Riebel2010} shows that the slope of the O-rich AGB PL relation is relatively insensitive to wavelength, whereas the C-rich relation shows significant variation.

Correcting for distance, the SMC stars are well-mixed with the LMC stars, showing comparable scatter about the best-fit line.  With many more samples than the O-rich, this provides more substantial evidence that C-rich AGB stars in the SMC follow the same PL relationship as those in the LMC.

\begin{deluxetable}{lccc}
\tabletypesize{\small}
\tablecaption{Period Luminosity Relationships for C-rich AGB stars Pulsating in the Fundamental Mode}
\tablewidth{0pt}
\tablehead{\colhead{PL Relation} & \colhead{N} & \colhead{scatter (mag)}}
\startdata
This work: \\
$[3.6] = -3.63(\pm 0.02) \log P + 19.11(\pm 0.05)$ & 272 & $0.24$ \\
$[4.5] = -3.84(\pm 0.02) \log P + 19.66(\pm 0.06)$ & 272 & $0.30$ \\
\citet{Riebel2010}: \\
$[3.6]=-3.77(\pm 0.05) \log P + 19.35(\pm 0.12)$ & 1813 & $0.251$ \\
$[4.5]=-3.56(\pm 0.05) \log P + 18.96(\pm 0.12)$ & 1816 & $0.265$
\enddata
\tablecomments{PL relations for LPVs classified as C-rich AGB stars.  The quoted scatter is the standard deviation of the residuals about the best-fit line.}
\label{tab:cagb_pl}
\end{deluxetable}

\subsubsection{Cepheid IR PL Relationships}\label{subsec:ceph_pl}
The well-known period-luminosity relationship of classical Cepheid variables (the Leavitt Law) is one of the most  important extragalactic distance indicators in astronomy.  Cepheid studies have generally focused on the optical, but \citet{McGonegal1982} pointed out that the near-IR offers several advantages for calibration of the Cepheid PL relation.  Interstellar reddening, which imposes an intrinsic scatter on magnitude determinations in the optical, is much smaller at \emph{Spitzer} wavelengths.  Also, Cepheid variability amplitudes are considerably smaller in the IR, and thus sparsely sampled lighcurves are good tracers of the mean brightness.   For these reasons, several recent surveys have focused on calibrating the Cepheid PL relation in the IR \citep{Scowcroft2011,Freedman2012,Ngeow2012}.

These studies have focused on small ($\sim$80) samples of Cepheids, with well sampled light curves.  SAGE-Var was designed as a more general variability survey, so what we lack in thorough coverage of the light-curve, we make up for in increased sample size.    Based on OGLE classification, we find 837 (1536) fundamental mode Cepheids in our LMC (SMC) data.    

We extract every source observed by SAGE-Var and classified as a fundamental mode Cepheid by the OGLE survey (Cepheids primarily varying in the first or higher overtone mode will by definition not lie on the primary PL relation).  We eliminate any source with only one valid observation during SAGE-Var, as this provides limited means to constrain the uncertainty on the mean flux of the source.  We calculate a simple mean magnitude for all our sources by averaging the fluxes of all SAGE-Var observations and then converting to magnitudes.  Only 4 of our sources are in common with \citet{Scowcroft2011}, and our mean magnitudes derived from randomly-phased observations are within 0.05 magnitudes of theirs.  We use the periods from the OGLE survey \citep{ogle_cep_lmc, ogle_cep_smc}.

\begin{figure}
\centerline{
\begin{tabular}{c}
\rotatebox{270}{\includegraphics*[width=0.65\linewidth]{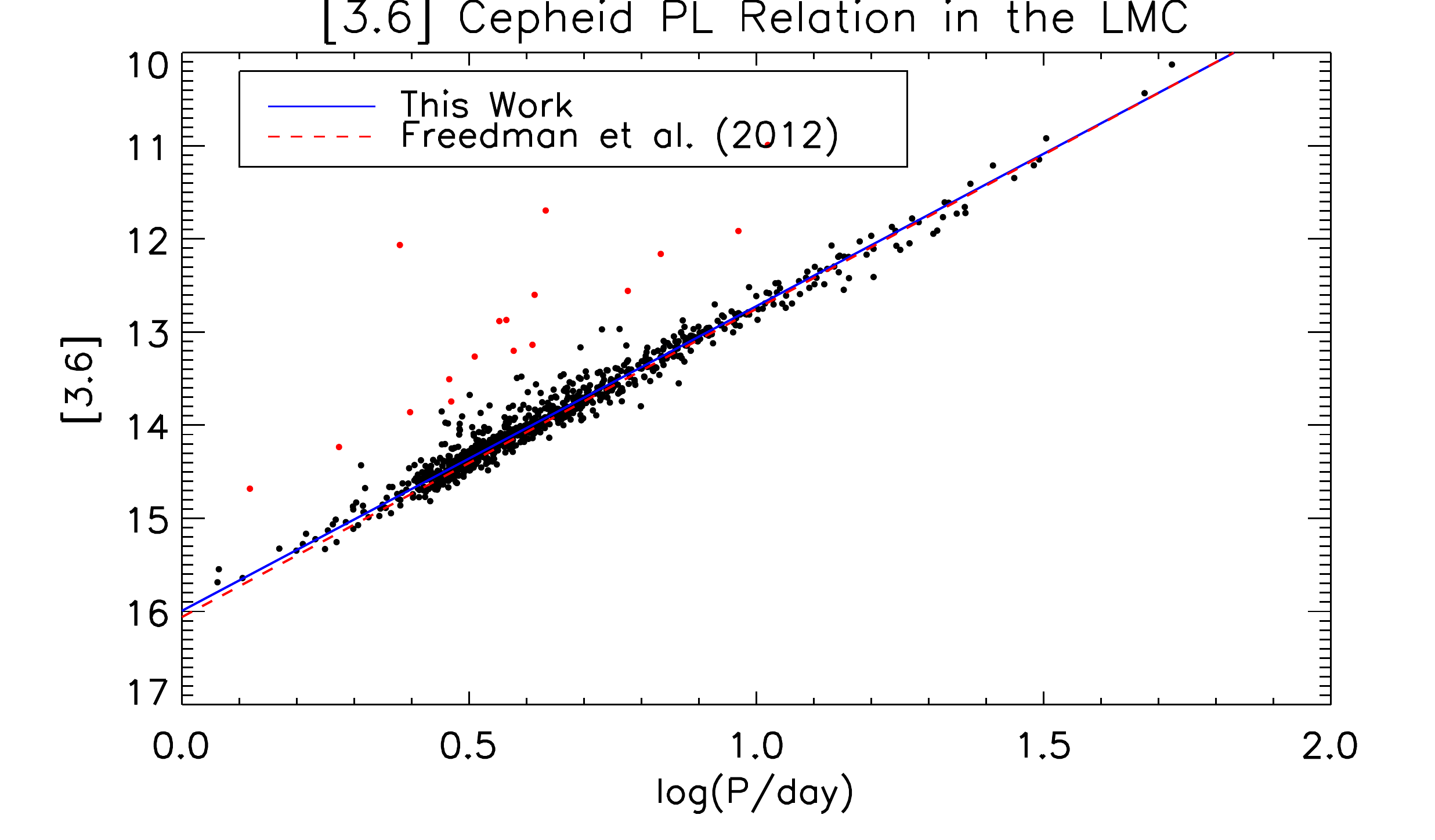}}\\
\rotatebox{270}{\includegraphics*[width=0.65\linewidth]{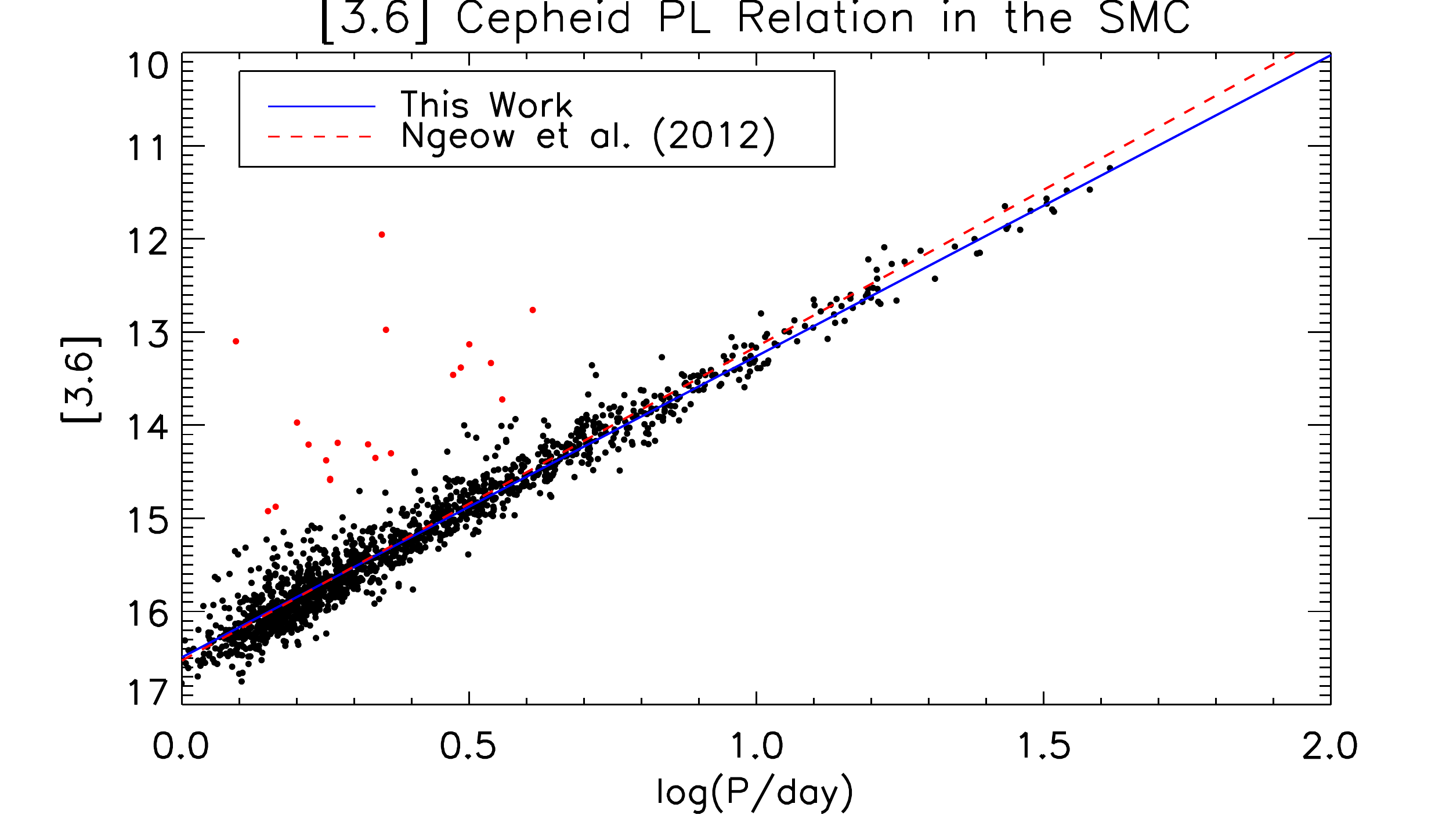}}
\end{tabular} }
\caption[Leavitt Law at 3.6\,\mic]{3.6\,\mic\ Leavitt Law for 811 (452) classical Cepheids in the LMC (SMC).  The LMC is shown in the top panel and the SMC in the bottom.  The fits are quantitatively detailed in Table~\ref{tab:ceph_pl}.  Stars shown in red have residuals to the fit greater than 3$\sigma$ and did not contribute to the fit.  The relation derived in this study is plotted in blue.  For the LMC (SMC) data, the relation determined by \citet{Freedman2012} \citep{Ngeow2012} is overplotted in red as a dashed line.}
\label{fig:ceph_pl_3}
\end{figure}

\begin{figure}
\centerline{
\begin{tabular}{c}
\rotatebox{270}{\includegraphics*[width=0.65\linewidth]{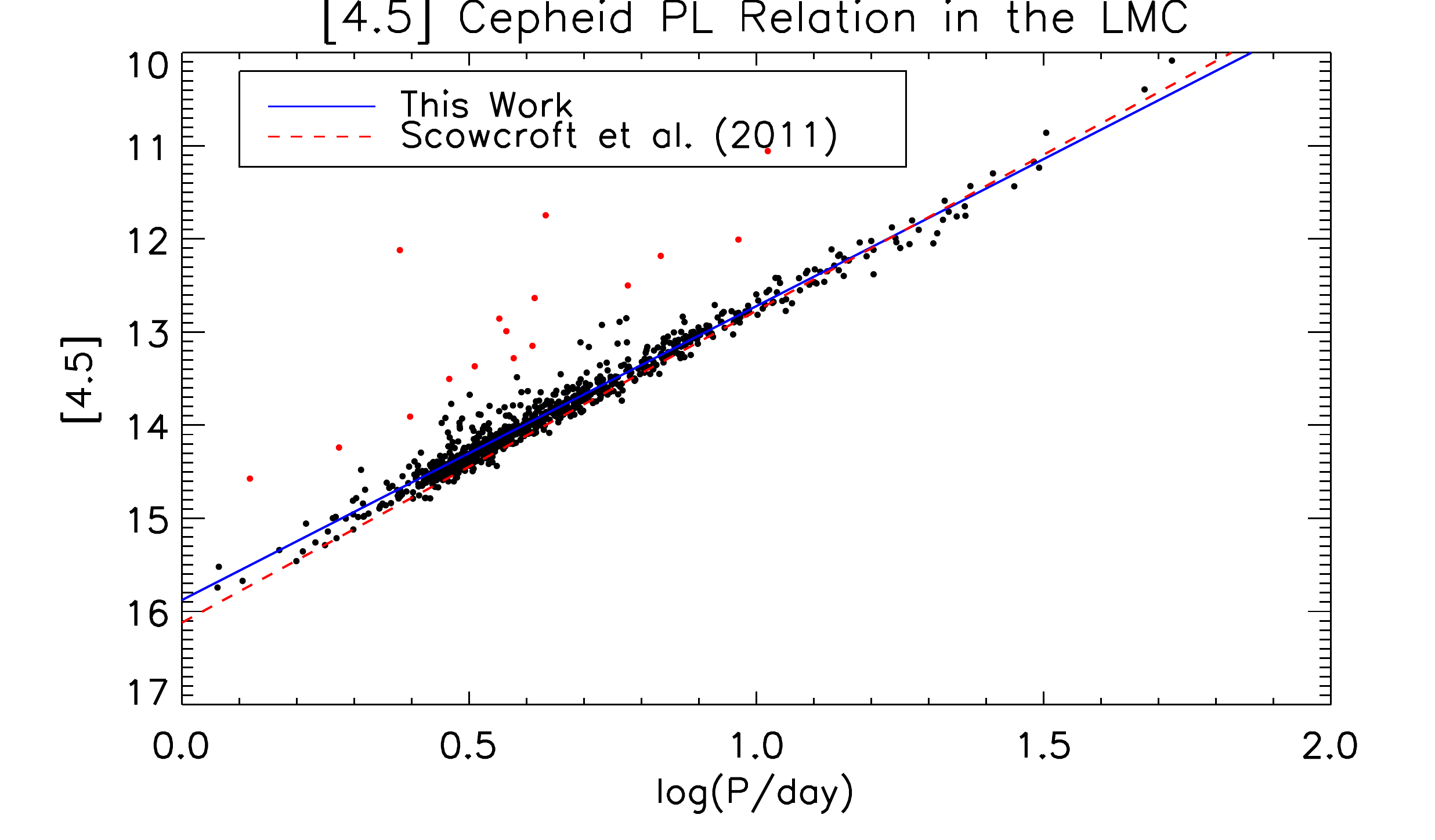}}\\
\rotatebox{270}{\includegraphics*[width=0.65\linewidth]{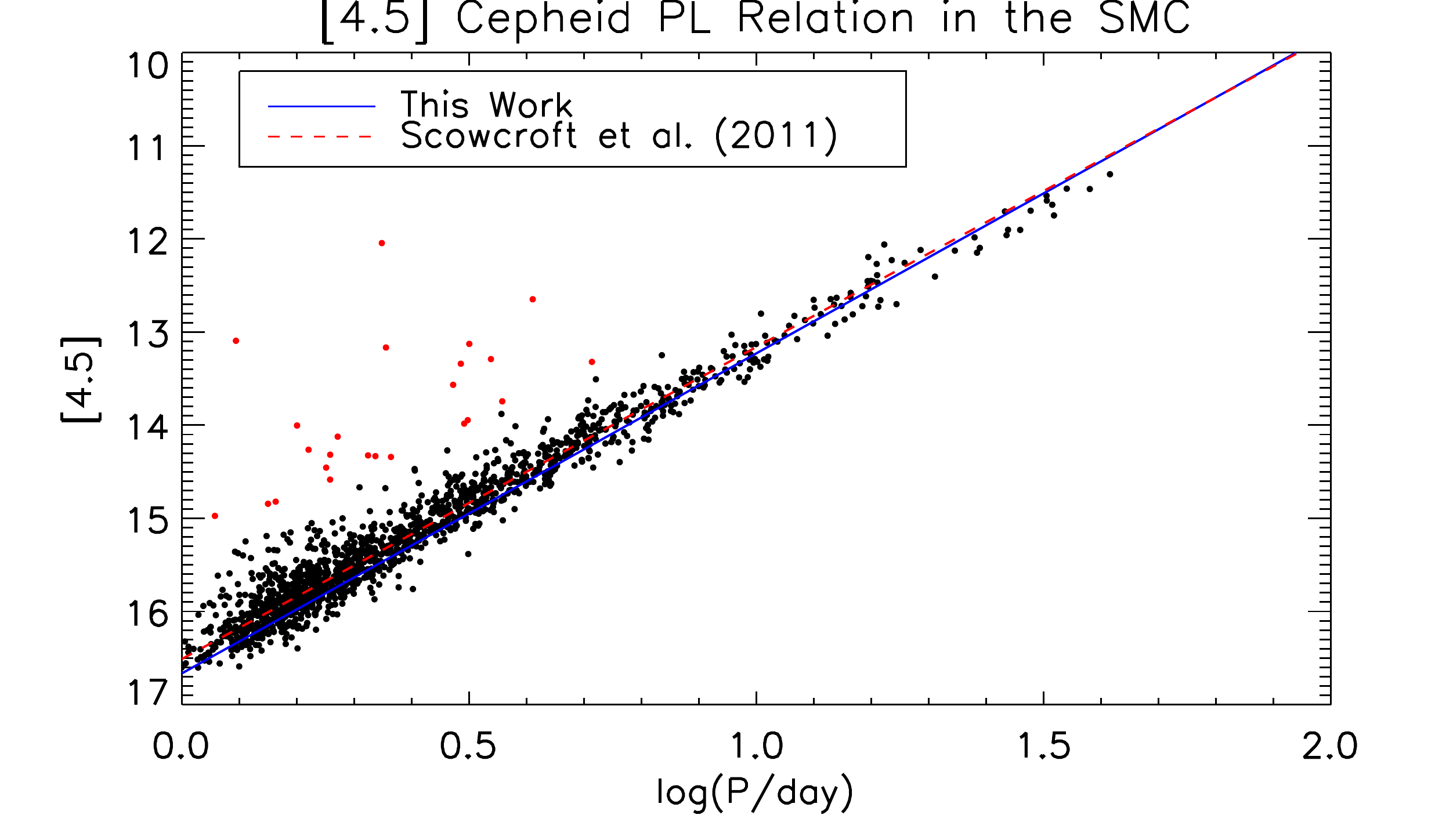}}
\end{tabular} }
\caption[Leavitt Law at 4.5\,\mic]{4.5\,\mic\ Leavitt Law for 811 (452) classical Cepheids in the LMC (SMC).  The LMC is shown in the top panel and the SMC in the bottom.  The fits are quantitatively detailed in Table~\ref{tab:ceph_pl}.  Stars shown in red have residuals to the fit greater than 3$\sigma$ and did not contribute to the fit.  The relation derived in this study is plotted in blue.  The relation determined by \citet{Scowcroft2011} using the original SAGE LMC data is overplotted in red as a dashed line.}
\label{fig:ceph_pl_4}
\end{figure}

The least squares linear fits (using 3$\sigma$ clipping) are given below in Table~\ref{tab:ceph_pl}.  The Leavitt Law at [3.6] derived from the  LMC sample is illustrated in Figure~\ref{fig:ceph_pl_3} and the relation at [4.5] is shown in Figure~\ref{fig:ceph_pl_4}.  Fits are of the form $y=A \log P + B$, with $y$ the mean magnitude of the source, and the period $P$ measured in days.  The scatter is defined as the standard deviation of the residuals to the fit.  We visually inspected all the sources clipped as part of the fitting process.   More than 90\% are obvious blending/confusion issues in the original data.  We find that when fitting the SMC data, using only sources with $\log P > 0.5$ provided a better visual fit to the data \citep[this method was also used by][]{Scowcroft2011}.  The LMC fits were robust to this decision, and no period selection criteria were applied.  The results given in Table~\ref{tab:ceph_pl} show the greater scatter about the SMC relationship compared with that in the LMC.  This is consistent with the much greater overall line-of-sight depth of the SMC compared with LMC seen in the early Cepheid work of \citet{Caldwell1986}, as well as in other stellar components.  This is probably due to gravitational interactions in the Magellanic system \citep[see][and references therein]{Scowcroft2015}.

 The slope of our [3.6] relations show very good agreement with those found by \citet{Freedman2012} in the LMC and by \citet{Ngeow2012} in the SMC.  In the [4.5] band, we overplot the relations determined by \citet{Scowcroft2011}, which also show good agreement with our own.  As discussed by \citet{Scowcroft2011} and \citet{Monson2012}, CO absorption in the [4.5] band renders this PL relation problematic for Cepheid distance determinations.  This effect is clearly seen in Table~\ref{tab:ceph_pl}, where the slopes of the PL relations at [3.6] in the the LMC and SMC agree well, while the slopes at [4.5] do not. This seems likely to be due to the different effects of CO absorption at different metallicities.

\begin{deluxetable}{lccc}
\tabletypesize{\scriptsize}
\tablecaption{Classical Cepheid Period Luminosity relationships at [3.6] and [4.5] in the Magellanic Clouds}
\tablewidth{0pt}
\tablehead{\colhead{PL Relation} & \colhead{N} & \colhead{scatter (mag)}}
\startdata
LMC: \\
$[3.6] = -3.271(\pm 0.004) \log P + 15.993(\pm 0.003)$ & 811 & $0.13$ \\
$[4.5]= -3.157(\pm 0.004) \log P + 15.877(\pm 0.002)$ & 820 & $0.13$ \\
SMC: \\
$[3.6]=-3.261(\pm 0.006) \log P + 16.511(\pm 0.004)$ & 452 & $0.18$ \\
$[4.5]=-3.437(\pm 0.004) \log P + 16.665(\pm 0.002)$ & 454 & $0.18$
\enddata
\tablecomments{PL relations for fundamental mode classical Cepheids in the LMC and SMC.  Fits were determined using a standard 3$\sigma$ clipping procedure.  The fits to the SMC data were determined by only considering Cepheids with $\log P > 0.5$.  The scatter about the fit is defined to be the standard deviation of the residuals.}
\label{tab:ceph_pl}
\end{deluxetable}

\section{Conclusions}\label{sec:conclusions}

We present the results from a 4-epoch unbiased IR survey of the central regions of the LMC and SMC.  We have produced full catalogs of our observations, consisting of \numlmc\ (\numsmc) objects in the LMC (SMC).  We have identified  \numlvars\ (\numsvars) objects in the LMC (SMC) as probable variables.  

We identify 10 (6) variable AGB candidates in the LMC (SMC) without well-determined variable periods from OGLE or MACHO (\S\,\ref{sec:new_lpvs}).  Most of these sources were flagged as potential variables in the WISE survey, and our independent measurement confirms that probability.  

Using mean magnitudes constructed from 6 epochs of observations, we investigate the PL relationship for AGB stars pulsating in the fundamental mode.  We find no significant reduction in the scatter about the best-fit relation compared to the results of \citet{Riebel2010}, indicating the scatter ($\sim$0.2 mag) might be intrinsic to the data (\S\,\ref{subsec:agb_pl}).  We find no evidence that the PL relationships of AGB stars in the SMC (Z$\sim$0.04) and in the LMC (Z$\sim$0.08) are different due to metallicity effects.

We present infrared PL relations for a sample of 811 (452) Cepheids in the LMC (SMC).  Cepheid amplitudes are small in the IR, and our random-cadence results compare very well with those of \citet{Scowcroft2011},  \citet{Freedman2012} and \citet{Ngeow2012} (\S\,\ref{subsec:ceph_pl}).

\section{Ackowledgements}
This publication made use of the VizieR data base operated by CDS, Strasbourg, France. 

MWF and PAW gratefully acknowledge the receipt of research grants from the National Research Foundation (NRF) of South Africa. 

Meixner acknowledges support from NASA NAG5-12595, and Spitzer contract NM0710076.

This work is based on observations made with the {\it Spitzer} Space Telescope, which is operated by the Jet Propulsion Laboratory, California Institute of Technology under a contract with NASA.

MLB is supported by the NASA Postdoctoral Program at the Goddard Space Flight Center, administered by ORAU through a contract with NASA.

SS acknowledges support from the National Science Council and the Ministry of Science and Technology in the form of grant MOST103-2112-M-001-033-.

\end{document}